\documentclass[a4paper,fleqn,usenatbib]{mnras}
\usepackage{newtxtext,newtxmath}

\usepackage[T1]{fontenc}
\usepackage{ae,aecompl}

\usepackage{graphicx}	
\usepackage{amsmath}	
\usepackage{amssymb}	

\newcommand{\openvirg}{``}
\newcommand{\closedvirg}{''$\ $}

\title[IR Spectral Fingerprint of Carbon Monoxide in Interstellar Water Ice Models]{IR Spectral Fingerprint of Carbon Monoxide in Interstellar Water Ice Models}

\author[L. Zamirri et al.]
{Lorenzo Zamirri$^{1,2}$,
Silvia Casassa$^{1,2}$,\thanks{e-mail:silvia.casassa@unito.it}
Albert Rimola$^{3}$,
Mireia Segado-Centellas$^{4}$,
\newauthor
Cecilia Ceccarelli$^{5,6}$, 
and Piero Ugliengo$^{1,2}$,\thanks{e-mail:piero.ugliengo@unito.it}
\\
\\
$^1$Dipartimento di Chimica, Universit\`a degli Studi di Torino, via P. Giuria 7, I-10125, Torino, Italy\\
$^2$Nanostructured Interfaces and Surfaces (NIS) Centre, Universit\`a degli Studi di Torino, via P. Giuria 7, I-10125, Torino, Italy\\
$^3$Departament de Qu\'imica, Universitat Aut\`onoma de Barcelona, E-08193 Bellaterra, Spain\\
$^4$Institut Catal\`a d'Investigaci\'o Qu\'imica (ICIQ), Avinguda Pa\"ios Catalans 16, E-43007, Tarragona, Spain\\
$^5$Institut de Plan\'etologie et d'Astrophysique de Grenoble (IPAG), rue de la Piscine, F-38041, Grenoble, France\\
$^6$INAF-Osservatorio Astrofisico di Arcetri, largo E. Fermi 5, I-50125, Firenze, Italy
}
\date{Accepted XXX. Received YYY; in original form ZZZ}

\pubyear{2018}

\begin{document}
\label{firstpage}
\pagerange{\pageref{firstpage}--\pageref{lastpage}}
\maketitle

\begin{abstract}
Carbon monoxide (CO) is the second most abundant molecule in the gas-phase of the interstellar medium. In dense molecular clouds, it is also present in the solid-phase as a constituent of the mixed water-dominated ices covering dust grains. Its presence in the solid-phase is inferred from its infrared (IR) signals. In experimental observations of solid CO/water mixed samples, its IR frequency splits into two components, giving rise to a blue- and a redshifted band. However, in astronomical observations, the former has never been observed. Several attempts have been carried out to explain this peculiar behaviour, but the question still remains open. In this work, we resorted to pure quantum mechanical simulations in order to shed some light on this problem. We adopted different periodic models simulating the CO/H$_2$O ice system, such as single and multiple CO adsorption on water ice surfaces, CO entrapped into water cages and proper CO:H$_2$O mixed ices. We also simulated pure solid CO. The detailed analysis of our data revealed how the quadrupolar character of CO and the dispersive forces with water ice determine the energetic of the CO/H$_2$O ice interaction, as well as the CO spectroscopic behaviour. Our data suggest that the blueshifted peak can be assigned to CO interacting {\it via} the C atom with dangling H atoms of the water ice, while the redshifted one can actually be the result of CO involved in different reciprocal interactions with the water matrix. We also provide a possible explanation for the lack of the blueshifted peak in astronomical spectra. Our aim is not to provide a full account of the various interstellar ices, but rather to elucidate the sensitivity of the CO spectral features to different water ice environments.  
\end{abstract}

\begin{keywords}
astrochemistry, ISM: lines and bands, ISM: molecules, molecular processes, methods: numerical, techniques: spectroscopic
\end{keywords}




\section{Introduction}\label{sec:Intro}
In cold ($\lesssim$ 100 K) molecular clouds, carbon monoxide (CO) is the second most abundant gaseous molecule (after hydrogen, H$_2$). In addition, given its spectroscopic and chemical properties, CO is widely used as a proxy of H$_2$ of molecular gas in both galactic and extragalactic sources. However, in very cold ($\lesssim$ 25 K) and very dense ($\gtrsim 10^4$ cm$^{-3}$) gas, CO molecules freeze-out onto the interstellar grain surfaces, in the so-called grain mantles, and disappear from the gas \citep{Caselli_CO_depl}. The effect can be extreme in particularly dense regions, as prestellar cores and protoplanetary disks, where more than 90\% of CO is likely in the solid- rather than gas-phase \citep{Bacmann_2002, Favre_2013}. Therefore, solid CO may be the major reservoir of elemental carbon (not locked into carbonaceous grains) in a large fraction of the cold interstellar medium (ISM).\\
The way how CO freezes, namely in what structure (whether trapped in a water-rich matrix or in pure CO layers), affects the temperature at which it returns into the gas-phase. In turn, this has important consequences on a variety of situations; for example, on the molecular deuteration of water and trace species  \citep{Ceccarelli_deut_2014}, on the formation of methanol and more complex organic molecules in the protostellar phase \citep{Garrod_2006}, or on the composition of the gaseous giant planets \citep{Oberg_2016, Madhusudhan_2017}.\\
Therefore, understanding the molecular structure of the interstellar solid CO is of great importance. This can only be obtained {\it via} observations of the 4.647-4.682 $\mu$m (2152-2136 cm$^{-1}$) $^{12}$CO and 4.780 $\mu$m (2092 cm$^{-1}$) $^{13}$CO absorption features towards astronomical sources and their comparison with ``expected'' solid CO spectra, based on laboratory and/or theoretical works.\\
From a theoretical point of view, grain mantles grow in molecular clouds and in the denser fragments of them, called prestellar cores. Water ice forms {\it via} hydrogenation of oxygen atoms which land on the grain surfaces \citep{Dulieu_2010, Oba_2012} whereas CO ice results directly from the sticking of gaseous CO onto the grains. Observations and models clearly indicate that water ice is the most abundant component of the grain mantles and that it is formed before the CO ice \citep{ARAA_Boogert_2015, Taquet_2013}. Therefore, in principle CO molecules can be trapped into a water-rich matrix or in pure-CO layers. In addition, UV photons and cosmic rays are believed to affect the iced mantle structure \citep{AA_Oberg_2009}. As a result of these several different phenomena, ice seems to be present in two rather different phases: $i$) as a highly porous amorphous film containing several apolar molecules and characterized by rather large cavities \citep{AA_AlHalabi_2004} and $ii$) as a layered and more ordered highly hydrogenated phase distinguished by a significant concentration of dangling groups, {\it i.e.} water molecules at a surface or discontinuity that can not complete their tetrahedrally hydrogen bonded network with their neighbouring molecules \citep{Collings_2003a}.\\
The CO/H$_2$O interplay takes place in several different and peculiar environments, including low pressure and temperature, and is expected to include several competitive aspects such as surface-surface interactions, CO adsorption, migration into ice pores \citep{JPhysChem_Devlin_1992} and CO trapping into the ice matrices \citep{AstroPL_Schmitt_1989, AstroJ_Sandford_1988}. Numerous experiments have addressed different aspects of the pure-H$_2$O and of the CO-containing ices, yielding a great number of data that prompt multiple and sometimes controversial issues. In particular, infrared (IR) spectral analysis is the main experimental instrument to investigate the physical-chemistry of CO interacting with H$_2$O ice
under interstellar-like conditions. However, it is sometimes difficult to orient in the plethora of experimental peaks, often faint and concentrated in a small portion of the spectrum. In this regard, quantum mechanical simulations can represent fundamental tools for deriving important structural and atomistic information.
The present manuscript is intended to calculate, within the same theoretical first principle approach, the IR fingerprints of several different atomistic structural models for the CO/H$_2$O ice interaction in order to facilitate the comparison among the various spectroscopic signals and to provide tentative molecular models for each of them in a coherent and robust framework.\\
The paper is organized as follows. In Section \ref{sec:revi-prev-stud}, we briefly review the situation about the laboratory and theoretical works on the IR absorption bands observed in the astronomical sources and the open questions about the interpretation of these observations. In Section \ref{sec:comp-data} we resume the adopted computational methods for static and dynamic calculations. In Section \ref{sec:results}, we present and discuss the results concerning the different CO/H$_2$O ice systems we modelled in the light of the most recent experimental findings and theoretical analysis. General hypothesis on some controversial issues are suggested at each subsection. Finally, in Section \ref{sec:conclusions}, all results are briefly summarized and some perspectives are also drawn.
\section{Brief review of previous studies}\label{sec:revi-prev-stud}
In IR observations, gas-phase CO is detected at 2143 cm$^{-1}$ (4.666 $\mu$m) due to its C$-$O stretching mode \citep{JChemEdu_Mina_1996}. Therefore, the analysis of the ISM spectra focuses on the splitting and shifts of this signal towards lower (red) or higher (blue) wavenumbers. Thus, before any investigation, an overview of the detected IR signals for CO/H$_2$O systems along with their interpretations is done here.\\
As already mentioned in the Introduction, two rather different ice structures are supposed to be present in the ISM, referred to as the apolar mixed (\textsc{Ap}) and the layered polar (\textsc{Pp}) phase. In order to reproduce both of them, as well as their interaction with carbon monoxide, different experimental settings and computational simulations have been performed over the years.\\
CO co-deposited in water matrices at low temperatures, as a model of \textsc{Ap}, presents a very peculiar band profile consisting of two main broad features: $i$) an intense peak at 2138 cm$^{-1}$ (4.677 $\mu$m), redshifted with respect to the gas-phase and indicating a decreasing of the bond strength (and a corresponding increasing of the C$-$O bond length), and $ii$) a secondary, less intense, blueshifted peak at 2152 cm$^{-1}$ (4.647 $\mu$m) corresponding to a shortening/increasing of the bond length/strength. This two-peaks profile has been noted and discussed by many authors.\\
According to \citet{AstroJ_Sandford_1988} and \citet{AstroJ_Jenni_1995}, the 2138 cm$^{-1}$ band is due to substitutional CO molecules, {\it i.e.} CO replacing H$_2$O lattice molecules, while the 2152 cm$^{-1}$ band is caused by interstitial CO, {\it i.e.}  CO inside H$_2$O cages. An alternative interpretation attributes the 2152 cm$^{-1}$ peak to CO molecules interacting with the OH dangling groups \citep{JPhysChem_Devlin_1992, Collings_2003a}, hereafter referred to as \textsc{dH} sites, while the 2138 cm$^{-1}$ is assigned to CO diffused into micropores \citep{JPhysChemA_Palumbo_1997}.\\
Interstellar H$_2$O ice is believed to be predominantly amorphous, although its specific morphology remains an open question when referring to its porosity. To investigate on the possibility of CO migrating into ice lattice, Al-Halabi and coworkers \citep{AA_AlHalabi_2004} have studied the interaction of CO with amorphous ice by means of classical trajectory calculations and have concluded that CO neither penetrate into the ice slab matrix nor diffuse into the surface valley. The eventuality of CO diffusion appears as a temperature effect \citep{JPhysChemA_Palumbo_1997}: when ice is warmed up to 30 K, CO mainly desorbs but few molecules are trapped in the H$_2$O porous structure giving rise to a narrowing of the band at 2138 cm$^{-1}$ related to CO in water matrix. At temperatures higher than 80 K, the disappearance of the peak at 2152 cm$^{-1}$ indicates the break of the \textsc{dH}$\cdot\cdot$CO interaction. Interestingly, the annealing process of CO-rich ices produces a family of new adsorption features, at 2143 cm$^{-1}$, that have been tentatively associated with the formation of clathrate-like structures \citep{AstroJ_Sandford_1988}.\\
The CO-layered ice binary system \textsc{Pp}, although prepared following different experimental procedures and theoretical models, presents almost the same spectroscopic pattern. At low CO coverage, it can be seen as a gas/surface interaction. When CO concentration increases, a CO/H$_2$O interface appears and a layered CO structure starts to grow. To characterize this system, \citet{JPC_Allouche_1998} have performed a combined experimental/computational study of CO adsorption on ice surfaces. They measured an adsorption enthalpy of about 10 kJ mol$^{-1}$ and unambiguously assigned the peak at 2152 cm$^{-1}$ to CO interacting with \textsc{dH}. In the context of this tendency toward a long range crystalline order, the peak at 2143 cm$^{-1}$ has been assigned to the adsorption of linearly polarized light in the CO ice phase, which starts to crystallize as the temperature increases \citep{AA_Pontoppidan_2003}. Finally, two combined effects are supposed to give rise to the 2139-2136 cm$^{-1}$ features: CO/CO interactions inside the rising CO multilayer as well as interactions involving dangling O atoms (\textsc{dO}) of the exposed water molecules occurring at the CO/H$_2$O interface \citep{JPL_Manca_2000, SurfSci_Martin_2002}. This band narrows as the temperature increases, due to the progressive CO desorption \citep{JChemPhys_AlHalabi_2004}.\\
While nowadays there is a currently general agreement as regard the peak at 2152 cm$^{-1}$, present both in the apolar mixture and the hydrogenated ice surfaces, and assigned to CO adsorbed on \textsc{dH} sites through its carbon atom, the nature of the 2143 and 2139-2136 cm$^{-1}$ peaks remains still controversial. Rather surprising, the 2152 cm$^{-1}$ signal has not been found in any of the interstellar spectra \citep{Collings_2003a, JChemPhys_AlHalabi_2004} while signals around 2143 cm$^{-1}$ appear as a prominent shoulder of the main peak as well as the redshifted features between 2139-2136 cm$^{-1}$.\\
In the following, we will explore at the same level of accuracy many different configurations of the CO/H$_2$O ice binary system and we will analyze the IR fingerprint for each of them.
\section{Computational Details}\label{sec:comp-data}
\subsection{Static calculations}\label{sec:QM}
We used the {\it ab initio} \textsc{CRYSTAL17} program \citep{CRYSTAL17_article,CRYSTAL17_manual} for all the calculations on periodic models. This code implements the Hartree-Fock and Kohn-Sham self-consistent field methods for the study of periodic and molecular systems \citep{Pisani_book} adopting Gaussian type orbital (GTO) functions as basis sets and allowing full geometry optimizations (both internal coordinates and cell size \citet{CRYSTAL_optgeom}) as well as characterization of the electronic structure and vibrational properties of molecules, surfaces and crystals. The GTO basis set we used for all calculations consists in a triple-$\zeta$ quality basis set developed by Ahlrichs and coworkers where polarization functions have been added (VTZ*)
\citep{Ahlrichs_basiset}. For more details about the basis set, please see Table A1 of the Appendix A file available online.\\
The pure GGA Perdew-Burke-Ernzerhof (PBE-D2) exchange correlation functional \citep{PBE} has been adopted for all calculations, in which D2 is the Grimme's correction for dispersive interactions \citep{GrimmeD2} as modified for molecular crystals \citep{Grimme_Crystal}.\\
In order to have an accurate sampling of the potential energy surface, we used a very large integration grid consisting in 99 radial points and 1454 angular points. The Hamiltonian matrix has been diagonalized \citep{Monkhorst} in a set of reciprocal {\it k}-points ranging from 14 to 260 as a function of the reciprocal lattice symmetry and cell dimension. For more details about the grid and the {\it k}-points, please see Table A2 of the Appendix A. Following previous experience \citep{Tosoni_2005}, and considering the very weak coupling between the CO stretching mode and other inter-molecular ones, the complete phonon frequency calculations are restricted to the Hessian matrix of the CO molecule(s) only as a fragment of the whole system. For all considered cases, we checked that the computed harmonic wavenumbers were all real, ensuring that the periodic models are all minima on the PBE-D2 potential energy surface, at least within the considered ensemble of CO molecule(s) \citep{Freq_Crystal}. The infrared intensity for each normal mode was obtained by computing the dipole moment variation along the normal mode, adopting the Berry phase method \citep{IR_freq, Berry_phase_crystal}.\\
When Gaussian basis sets are used, as in CRYSTAL17, the basis set superposition error ($\textsc{BSSE}$) arises in the evaluation of the interaction energy \citep{BSSE}. One of the most common {\it a   posteriori} correction is represented by the \openvirg Counterpoise\closedvirg (CP) method, which exploits \openvirg ghost function\closedvirg \citep{BSSE_CP}. For all tested cases, we computed the BSSE-corrected interaction energy {\it per} adsorbed CO molecule $(\Delta E^{\textsc{CP}})$ of the CO-ice complexes according to Equation \eqref{eq_DE1}
\begin{equation}
\Delta E^{\textsc{CP}} =\frac{\Delta E^{*}+\delta E+E_L-\textsc{BSSE}}{n} 
\label{eq_DE1}
\end{equation}
where $\Delta E^{*}$ is the deformation-free interaction energy, $\delta E=\delta E_I + \delta E_{\textsc{CO}}$ is the total deformation energy of both the ice model and the CO molecules, $E_L$ is the lateral interaction among different replicas of the periodic models, $\textsc{BSSE}$ is the total $\textsc{BSSE}$ contribution and $n$ is the number of CO molecules in the complex.
To fully account for quantum mechanical effects, we also computed the interaction enthalpies $\Delta H$ at 0 K according to Equation \eqref{eq_DH}
\begin{equation}
\Delta H = \Delta E^{\textsc{CP}} + \Delta\textsc{ZPE}
\label{eq_DH}
\end{equation}
where $\Delta\textsc{ZPE}$ is the difference (again {\it per} adsorbed CO molecule) between the zero-point energy ($\textsc{ZPE}$) of the $n$ CO molecules in the complexes (considered as a fragment of the whole structure) and $n$ times the $\textsc{ZPE}$ for the isolated CO, unless otherwise specified. For a more detailed discussion of Equation \ref{eq_DE1} and of the assumptions underlying Equation \ref{eq_DH}, please see the \openvirg Computational details\closedvirg section of the Appendix A.\\
We scaled all the harmonic computed CO stretching wavenumbers with a proper scaling factor defined as the ratio between the experimental (2143 cm$^{-1}$, \citet{JChemEdu_Mina_1996}) and our calculated (2118 cm$^{-1}$) value for the isolated CO molecule. This allows to partially recover for the intrinsic errors associated with the computational level we adopted and also to account for anharmonic effects.\\
The IR simulated spectra are defined as linear superpositions of Lorentzian functions centred at the different computed wavenumbers and weighted for the corresponding IR intensities. We choose the two arbitrary values of 3 and 15 cm$^{-1}$ as Full Width at Half Maximum (FWHM) to account for the natural width of the experimental and astronomical measures.\\
When dealing with 2D periodic slab models, we evaluated the relative stability of a given slab model compared to the corresponding bulk through the calculation of the {\it surface energy} $E_S$ according to Equation \eqref{eq_ES}
\begin{equation}
E_S=\frac{E_{slab}-mE_{bulk}}{2A}
\label{eq_ES}
\end{equation}
where $E_{slab}$ and $E_{bulk}$ are the total PBE-D2 energies of the slab and bulk model, respectively, $m$ is the ratio between the slab and bulk stoichiometries and $A$ is the slab unit area. The factor 2 accounts for both the top and bottom ends of the slab models.
\subsection{Dynamic simulations}\label{sec:MD}
We performed {\it ab initio} molecular dynamics (AIMD) simulations using the CP2K program which employs Gaussian-type and plane wave basis sets for the wavefunction and the electron density, respectively \citep{cp2k_soft}. We run AIMD simulations in order to sample the complex conformational space with the aim to amorphize CO/H$_2$O ice mixtures. All AIMD simulations are performed at the PBE level, enriched by the D3 {\it a posteriori} Grimme correction for dispersive interactions \citep{GrimmeD3}. A molecular optimized double-$\zeta$ basis set was applied to all atoms together with the corresponding Goedecker-Teter-Hutter (GTH) pseudopotential \citep{CompPhysComm_VandeVondele_2005}. The parameters used for the AIMD simulations are: 0.5 fs for the time-step, 15 ps for the production time, Nose-Hoover thermostat at 300 K and 1 bar for the pressure.

\section{Results and discussion}\label{sec:results}
In this section, we will discuss the structural, energetic and spectroscopic features of the different CO/H$_2$O ice systems we simulated in this work.
%
\subsection{Ice models}\label{sec:icemodesls}
We are well aware that interstellar ices are likely to be highly amorphous and porous. However, simulating amorphous systems is a non-trivial and highly subjective operation from a computational point of view. 
Keeping this in mind, we adopted two different 3D periodic models of ice to simulate crystalline ice phases, stable at low temperature and pressure ({\it i.e.} in ISM-like conditions): $i$) the ordinary hexagonal ice Ih, and $ii$) the orthorhombic proton-ordered polar ice XI (space group {$Cmc2_1$}) or \textsc{C-ice}. In turn, we modelled Ih ice with periodic (long-range ordered) structures belonging to the $ Pna2_1$ space group, usually referred as \textsc{P-ice}, with a unit cell providing a variety of configurations to describe the local proton disorder \citep{CPL_Casassa_1996,JCP_Casassa_2009}.\\
Specifically, we selected \textsc{P-ice} to model ice surfaces, by means of 2D periodic slabs, because it has local disorder, yet forms a simple (001) and (010) surfaces with no dipole component perpendicular to them, and essentially no dipole component parallel to them either \citep{CPL_Casassa_1996, JCP_Casassa_2002}.\\
Bulk \textsc{P-ice} was also used to simulate cages (as a single and 2x2x2 supercell) as well as to simulate a proper mixed ice system (1x1x2 supercell). Cages were also simulated by means of clathrate structures (namely, \textsc{sI} and \textsc{sIII}) and of a 2x2x2 \textsc{C-ice} supercell.
\subsection{Summary of CO/H$_2$O ice models}\label{sec:co_water_modesls}
The CO/H$_2$O ice interaction depends on the electronic properties of the CO molecule, such as its electrostatic potential (ESP) and its dipole and quadrupole moment. In Section \ref{sec:pureco} we provide a detailed analysis of these properties, together with those for crystalline solid CO systems. The latter can be representative of situations where \openvirg solid CO islands\closedvirg grow, under interstellar conditions, upon the dust grain mantles.\\
In Section \ref{sec:onpice} we describe the single adsorption of CO on exposed \textsc{dH} sites of two selected \textsc{P-ice} surfaces (\textsc{dH}$\cdot\cdot$CO/OC models). These models can be seen as references for the pure CO/H$_2$O ice interaction, since no other factors such as CO/CO repulsion occur. However, they are probably not proper representative of dirty ices grown under interstellar conditions because of the very low CO:H$_2$O molecular ratio.\\
To improve our description, in Section \ref{sec:inter} we deal with multiple CO adsorption on the (001) \textsc{P-ice} surface. These models can be representative of the formation of a solid-H$_2$O/solid-CO interface.\\
Since interstellar ices are expected to be highly porous \citep{AA_AlHalabi_2004}, in Section \ref{sec:cages} we explore the encagement of one or more CO molecules entrapped inside 3D periodic models of water cages of different size and shape, from the quite large cavities in clathrate named as \textsc{sI} and \textsc{sIII} to the very small ones in \textsc{P-} and \textsc{C-ice} bulks. For a complete description, we developed two 2D slab models where one CO is entrapped within \openvirg surface water cages'', {\it i.e.} cages formed by removing some water molecules at the surface of the water ice model (\openvirg expulsion\closedvirg models). We also studied CO substituting a H$_2$O molecule in the ice bulk (\openvirg Sub$_{\textsc{CO}}$\closedvirg model): calculations revealed, however, a restructuring of the CO surrounding water molecule in a clathrate-like fashion.\\
Finally, in Section \ref{sec:dirty} we simulate the spectroscopic and energetic properties of a mixed CO/H$_2$O ice system with a CO:H$_2$O molecular ratio close to the interstellar relative abundances in the solid phase.
\subsection{Pure CO}\label{sec:pureco}
\begin{figure*}
\includegraphics[width=17cm]{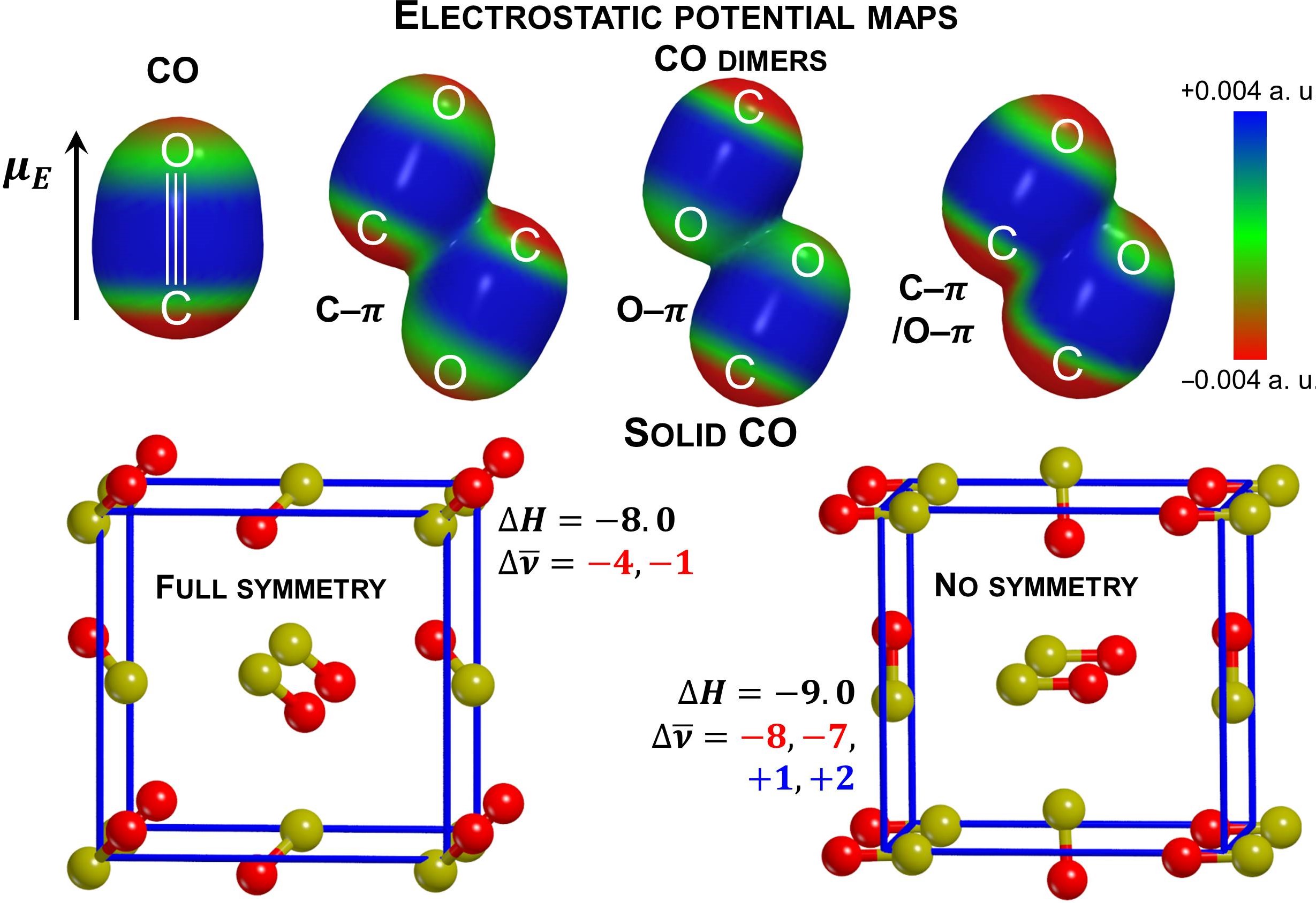}
\caption{Top: molecular electrostatic potential maps superimposed to the electron densities for the single CO molecule and the three energy minima of the CO dimer. The iso-density value for the electron density is set equal to 10$^{-5}$ a. u. $\boldsymbol{\mu_E}$ is the CO electric dipole moment vector. Bottom: optimized structures for solid $\alpha-$CO with $P{2_13}$ (left) and $P1$ (right) symmetry. Sublimation enthalpies {\it per} CO $\Delta H$ in kJ mol$^{-1}$, vibrational shifts $\Delta \bar{\nu}$ in cm$^{-1}$.}
\label{fig:pureco}
\end{figure*}

Gas-phase carbon monoxide possesses a small dipole moment $\boldsymbol{\mu_E}$ of 0.122 D oriented as C($\delta^-$)$\rightarrow$O($\delta^+$) \citep{JMSpec_Muenter_1975}. It is well known \citep{JCP_Scuseria_1991} that Hartee-Fock and several density functional theory (DFT) methods fail in the prediction of both the absolute value and the orientation of  the CO dipole. However, within our adopted computational scheme, CO dipole moment has the correct orientation, even if the module is slightly overestimated (0.196 D). From an electrostatic point of view, CO must be considered as a quadrupolar molecule due to its large quadrupole moment \citep{MP_Graham_1998}. In a recent work \citep{ESC_Zamirri_2017} we suggested that the quadrupole moment plays a fundamental role in determining the energetic and spectroscopic features of CO adsorbed on selected surfaces of crystalline forsterite as a model of the core dust grain. This quantity, not directly computed by the CRYSTAL code, can be eventually inferred by mapping the molecular ESP superimposed to the electron density. As it is shown in Figure \ref{fig:pureco}, the quadrupolar distribution of electric charges for gas-phase CO is clearly visible from its ESP map. The value of the {\it zz} component of the quadrupole, computed at the same level of theory of CRYSTAL17 calculations by means of the Gaussian09 code \citep{Gaussian09}, is  $-$1.4942 D \AA \ to be compared with the experimental value of $-$2.63 D \AA \ \citep{JChemPhys_Chetty_2011}.\\
In order to underline the importance of quadrupolar effects, we run very simple molecular calculations on different CO dimers. In 
Figure \ref{fig:pureco}, we report the structures corresponding to the only three energy minima we found. Regardless the initial geometry, in the final optimized ones the two CO molecules assumed an anti-parallel configuration, where the electric negative poles of one molecule (C or O atoms) are interacting with the \openvirg positive belt region\closedvirg of the other molecule, resembling the slipped pair in the benzene dimer, entirely dominated by quadrupole-quadrupole interactions. Initial geometries like {\it T-} or {\it L-shaped} ones results in saddle points of first and/or second order. The structures for all tested cases, as well as their spectroscopic features, are reported in the \openvirg Dipolar and quadrupolar interactions\closedvirg section of the Appendix A (Figures A13-A19 and Table A3). The C$-\pi$ interaction results to be almost 50\% more stable than the other two (O$-\pi$ and C$-\pi$/O$-\pi$), see $\Delta H$ column of Table \ref{tab:dimers}. Our results, although in contrast with those obtained in a previous work \citep{Collings_2014} at a lower level of theory, are in complete agreement with the fact that the C atoms represents the negative pole of the CO-dipole.\\
\begin{table}
\caption{Structural, energetic and vibrational features for the three CO dimers of Figure \ref{fig:pureco}. Interaction energies $\Delta E^{\textsc{CP}}$ and enthalpies $\Delta H$ {\it per} CO molecule in kJ mol$^{-1}$, wavenumbers $\bar{\nu}$ and shifts $\Delta \bar{\nu}$ with respect to the gas-phase CO stretching in cm$^{-1}$, distances in \AA. $d_{\textsc{C-O}}$ is the C$-$O bond length while ``X$-$X'' stands for ``C$-$C'' (C$-\pi$), ``O$-$O'' (O$-\pi$) and ``C$-$O'' (C$-\pi$/O$-\pi$).}
\label{tab:dimers}%
\begin{tabular}{lrrrrr}
\hline
 & $\Delta E^{\textsc{CP}\,\#}$ & $\Delta H$  & $d_{\textsc{C-O}}$ & $d_{\textsc{X-X}}$ & $\bar{\nu}$ ($\Delta \bar{\nu}$)\\
\hline
C$-\pi$			& $-$1.4 ($-$0.7) & $-$0.7 & 1.1367 & 3.249 & 2143 (0) \\
O$-\pi$ 		& $-$0.9 ($-$0.3) & $-$0.5 & 1.1368 & 3.177 & 2142 ($-$1) \\
C$-\pi$/O$-\pi$ & $-$1.0 ($-$0.3) & $-$0.5 & 1.1368 & 3.311 & 2142 ($-$1) \\
\hline
\end{tabular}
\\
$^{\#}$: $\Delta E^{\textsc{CP}}$ without the dispersion contributions in parenthesis. 
\end{table}

The other extreme is represented by solid $\alpha-$CO cubic crystal structure \citep{ZPhys_Vegard_1930} $-$ space group $P{2_13}$ $-$ with four molecules {\it per} unit cell located on equivalent sites of symmetry $C_3$. In the equilibrium configuration, the centres of the molecules are at the corners and face-centres of a cube and the internuclear axes are oriented along the body diagonals, as in Figure \ref{fig:pureco}.\\
Our optimized structure has a sublimation enthalpy at 0 K {\it per} CO ($\Delta_{\rm subl}H$) of 8.0 kJ mol$^{-1}$ (computed according to Equation \eqref{eq_DH}, including all ZPE contributions in the solid phase), This value is in excellent agreement with the experimental one measured at 10 K \citep{JCP_Kohin_1960}, of 8.3 kJ mol$^{-1}$, as reported in Table \ref{tab:solidCO}.\\
\begin{table}
\caption{Comparison between experimental and our computed 
data for crystalline solid $\alpha$-CO ($P2_13$ symmetry). $a_0$ lattice parameter 
in \AA, $ \Delta H$ sublimation enthalpy in kJ mol$^{-1}$, 
MIS \openvirg mean internuclear separation\closedvirg (average of the C$-$O bond lengths) in \AA \, and $\bar{\nu}$ 
CO stretching wavenumber in cm$^{-1}$.}
\label{tab:solidCO}
\begin{tabular}{lrrr}
\hline
Quantity & \,\,\,\,\,\, Experimental & \,\,\,\,\,\,  Our calculations &\,\,\,\,\,\,  $\%$Error \\
\hline
$a_0$ & $^a$5.64& 5.54 & 1.8 \\
$\Delta H$ & $^b$8.3 & 8.0 & 3.8 \\
MIS & $^c$1.128& 1.136 & 0.7 \\
$\bar{\nu}$ & $^d$2138.1& 2142.0 & 0.2 \\
\hline
\end{tabular}
\\
$^a$: \citet{PRB_Hall_1976}.\\
$^b$: \citet{JCP_Kohin_1960}. Measured at 10 K.\\
$^c$: \citet{PRB_Hall_1976}.\\
$^d$: \citet{JChemPhys_Ewing_1961}.
\end{table}

As regards the spectral fingerprint, the experimental stretching has been measured \citep{JChemPhys_Ewing_1961} at 2138 cm$^{-1}$ and the adsorption signal of the minor isotope, the $^{13}$CO, appears at 2092 cm$^{-1}$ \citep{JChemPhys_Ewing_1961}.\\
Our computed (scaled) stretching wavenumbers for solid CO are 2138 (IR inactive, $A$ symmetry, full in-phase stretching) and 2142 cm$^{-1}$ (IR active, $T$ symmetry, out-of-phase stretching) whereas, if the symmetry constrains are released to get closer to a poly-crystalline model, both vibrational modes become active and slightly decrease in value.
The computed stretching wavenumber for the $^{13}$C isotopic substituted $\alpha-$CO is 2094 cm$^{-1}$.\\
These small differences can be explained by different factors. Firstly, experimental IR spectra for solid CO show a some broadening in the band, centred at 2138 cm$^{-1}$, as a consequence of the several different contributions in a poly-crystalline-like phase, whereas our computed values refer to a perfect single crystal. Secondly, when scaling all computed wavenumbers for the same factor (computed from gas-phase data) we are assuming that all the perturbing factors in gas-phase (anharmonicity, method-related errors) are implicitly preserved in the solid phase. However, the absolute differences are almost negligible and, within all these assumptions, we are able to reproduce the most important experimental aspect, {\it i.e.} the significant redshift of solid CO with respect to the gas-phase.\\
\begin{table}
\caption{IR signals, in cm$^{-1}$, for pure CO systems. Experimental
data were recorded at T $\approx$ 10 K.}
\label{tab:allCO} 
\begin{tabular}{lrr}
\hline
Phase & \,\,\,\,\,\,\,\,\,\,\,\,\,\,\, Experimental  & \,\,\,\,\,\,\,\,\,\,\,\,\,\,\, Our calculations \\
\hline
gas    &   $^a$2143 & 2143 \\
liquid &   $^b$2138 &  $-$  \\
bulk ($\alpha$ phase) & $^c$2138 & 2142 \\
bulk $\alpha-^{13}$CO & $^c$2092 & 2094 \\
bulk LO/TO & $^d$[2143.7, 2138.5] & [2147, 2142] \\
\hline
\end{tabular}
\\
$^a$: \citet{JChemEdu_Mina_1996}.\\
$^b$: \citet{JChemPhys_Ewing_1962}.\\
$^c$: \citet{JChemPhys_Ewing_1961}.\\
$^d$: \citet{AA_Pontoppidan_2003}.
\end{table}

Finally, to enlighten on the LO/TO splitting origin of the peak at 2143.7 cm$^{-1}$, as suggested by Pontoppidan et al. \citep{AA_Pontoppidan_2003, Collings_2003b}, we have evaluated the CO dielectric constant $\epsilon = 1.69$ and calculated the LO wavenumber. Again, the computed value is overestimated with respect to the experimental signal, see Table \ref{tab:allCO}, but the interval shift of $\approx$ 5 cm$^{-1}$ between the LO and TO modes is perfectly reproduced. On this basis, the LO/TO hypothesis appears consistent although it cannot be observed in astronomical spectra \citep{Collings_2003a} and Boogert and coworkers \citep{ARAA_Boogert_2015} have recently found this assignment less likely.
\subsection{Single CO adsorption on water ice surfaces}\label{sec:onpice}
\begin{figure*}
\centering
\includegraphics[width=17.0cm]{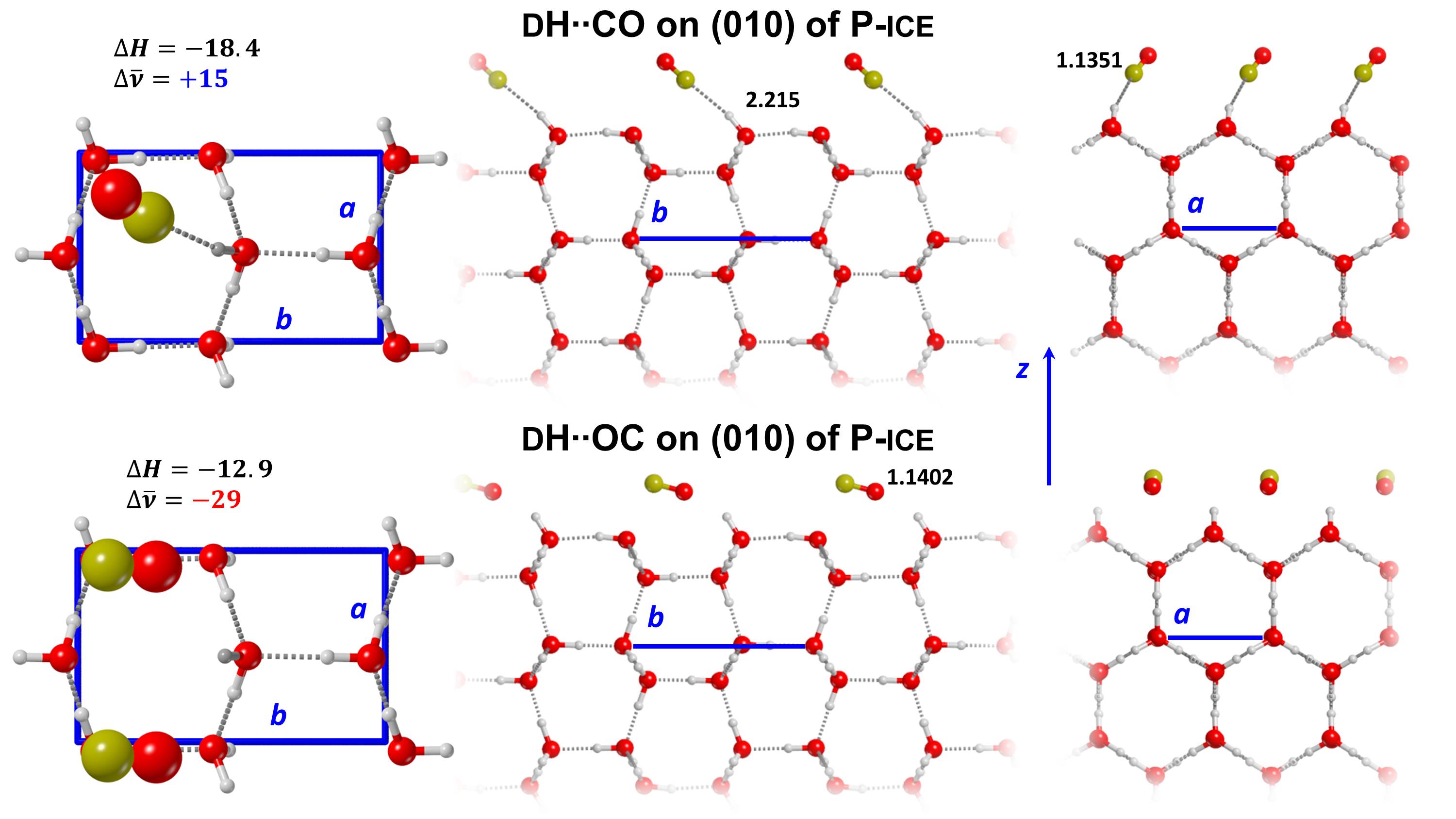}
\caption{Final optimized geometries for CO initially interacting with \textsc{dH} sites of the (010) \textsc{P-ice} surface {\it via} either C (\textsc{dH}$\cdot\cdot$CO) or O (\textsc{dH}$\cdot\cdot$OC) atom.
Top (left) and side views along $\boldsymbol{a}$ (centre) and $\boldsymbol{b}$ (right) periodic vectors. $\boldsymbol{z}$ is the direction perpendicular to the cut. In top views, C and O atoms of CO molecules as van der Waals spheres, \textsc{dH}s in dark grey.}
\label{fig:010}
\end{figure*}

\begin{figure*}
\centering
\includegraphics[width=17.0cm]{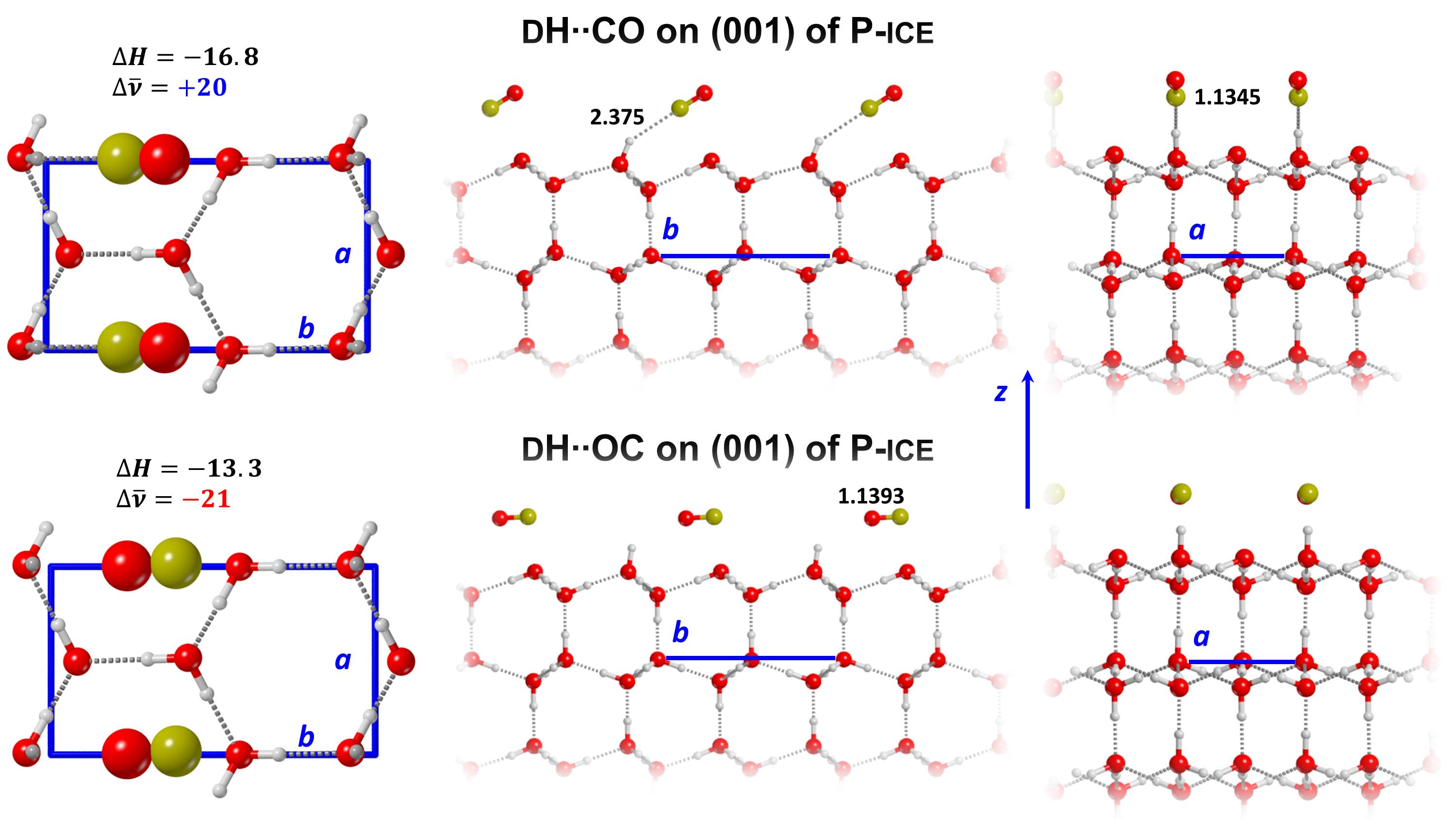}
\caption{Final optimized geometries for CO initially interacting with \textsc{dH} sites of the (001) \textsc{P-ice} surface {\it via} either C (\textsc{dH}$\cdot\cdot$CO) or O (\textsc{dH}$\cdot\cdot$OC) atom. Top (left) and side views along $\boldsymbol{a}$ (centre) and $\boldsymbol{b}$ (right) periodic vectors. $\boldsymbol{z}$ is the direction perpendicular to the cut. In top views, C and O atoms of CO molecules as van der Waals spheres, \textsc{dH}s in dark grey.}
\label{fig:001}
\end{figure*}

Single CO adsorption has been studied on two different ice surfaces cut out from \textsc{P-ice} bulk, corresponding to the (001)  and (010) faces. The 2D slabs, consisting of 12 water molecules {\it per} unit cell, as shown in Figure \ref{fig:010} and \ref{fig:001}, were fully optimized and results show that the (010) is slightly more stable than the (001) one (the former having a $E_S$ value 3.6 mJ m$^{-2}$ lower than that for the latter). Both surfaces expose in equal proportion dangling \textsc{dH} and \textsc{dO} atoms.\\
One CO molecule {\it per} unit cell was approached to the (010) and (001) \textsc{dH} surface sites through its C and O atoms. These configurations, reported in Figure \ref{fig:010} and \ref{fig:001} for the (010) and (001) cases and described in Table \ref{tab:CO_PICE}, will be referred in the following as \textsc{dH}$\cdot\cdot$CO and \textsc{dH}$\cdot\cdot$OC, respectively. On both surfaces, the initial \textsc{dH}$\cdot\cdot$CO interaction is preserved while the \textsc{dH}$\cdot\cdot$OC is lost during the optimization as dispersive and quadrupolar interactions between CO and the surface dominate over H-bonds. In all cases, CO quadrupolar moment exerts a crucial effect on both energetic and spectroscopic features. Irrespective of the surface, there are no evidences of a direct interaction involving \textsc{dO} sites, neither {\it via} the O atom (positive pole of the dipole) nor {\it via} the $\pi$ system (positive region of the quadrupole). This trend is also confirmed by calculations performed on a single CO interacting with the water-trimer, which represents the simplest system exposing both \textsc{dH} and \textsc{dO}. Structural, energetic and vibrational data for these systems are reported in the Appendix A (Figures A1-A12 and Table A3).\\
For \textsc{dH}$\cdot\cdot$CO cases, a definitive H-bond is formed between the C atom and the \textsc{dH} on both surfaces, resulting in a blueshifting of $\approx$ 15 cm$^{-1}$ of the CO stretching wavenumber with respect to the gas-phase and in a shortening of the C$-$O bond. These results are consistent with those obtained by Allouche and coworkers \citep{JPC_Allouche_1998, JPC_Allouche_2001} and further confirmed by \citet{JChemPhys_AlHalabi_2004} {\it via} molecular dynamics simulations. For \textsc{dH}$\cdot\cdot$OC cases, specific dispersive interactions cause an increasing of the C$-$O bond and a consequent redshift of $\approx-$20 cm$^{-1}$. Summarizing the results of this subsection we can state that: {\it i}) the \textsc{dH}$\cdot\cdot$CO interactions may explain the experimentally observed blueshifted peak at around 2152 cm$^{-1}$; {\it ii}) there is no interaction between CO molecules and \textsc{dO} sites; {\it iii}) quadrupolar and dispersive interactions play a fundamental role in determining the energetic and structural features of CO adsorption, particularly when CO interacts {\it via} its O atom with \textsc{dH} sites; {\it iv}) the resulting CO redshifted band does not provide a convincing explanation for the experimental signal at 2138 cm$^{-1}$.\\
\begin{table*}
\caption{Different configurations of CO interacting with perfect models of \textsc{P-ice} surfaces. Interaction energies $\Delta E^{\textsc{CP}}$ and enthalpies $\Delta H$ {\it per} CO molecule in kJ mol$^{-1}$ ($\Delta E^{\textsc{CP}}$ without the dispersion contribution in parenthesis), wavenumbers $\bar{\nu}$ and shifts $\Delta \bar{\nu}$ with respect to the gas-phase CO stretching in cm$^{-1}$, C$-$O distances $d_{\rm C-O}$ and H-bonds in \AA.}
\label{tab:CO_PICE}
\begin{tabular}{lrrrrrr}
\hline
 &\,\,\,\,\,\,\,\,\,\,\,\,\,\,\,\,\,\,\,\ & $\Delta E^{\textsc{CP}}$ (no disp.) \,  & \, $\Delta H$ \, & \,\, $\bar{\nu}$ ($\Delta \bar{\nu}$) \,\, &
\,\, $d_{\rm C-O}$ \,\, & \,\, H-bond \,\,  \\
\hline
 \multicolumn{7}{l}{CO adsorbed on \textsc{P-ice}} \\
 \multicolumn{7}{l}{(010) surface} \\
\textsc{dH}$\cdot \cdot$CO & & $-$21.0 ($-$12.9) & $-$18.4 & 2158 (+15) & 1.1351 & 2.215   \\   
\textsc{dH}$\cdot \cdot$OC & & $-$14.7 ($-$3.3)  & $-$12.9 & 2114 ($-$29) & 1.1402 & $-$   \\ 
\hline  
\multicolumn{7}{l}{(001) surface} \\
\textsc{dH}$\cdot\cdot$CO & & $-$19.1 ($-$9.9) & $-$16.8  & 2163 (+20)  & 1.1345 & 2.375  \\   
\textsc{dH}$\cdot\cdot$OC & & $-$15.3 ($-$5.4) & $-$13.3  & 2122 ($-$21)& 1.1393 & $-$     \\   
\hline
\multicolumn{7}{l}{CO/\textsc{P-ice} interfaces} \\ 
Interface 1 & & $-$5.1 ($-$2.9)    & $-$4.9 & 2133 ($-$10) & 1.1377 & $-$ \\
 & & & & 2140 ($-$3) & 1.1369 & $-$   \\
 & & & & 2147 (+4)   & 1.1360 & $-$   \\
 & & & & 2164 (+21)  & 1.1344 & 2.315 \\
Interface  2 & & $-$5.1 ($-$3.0)   & $-$5.0 &  2137 ($-$6) & 1.1372 & $-$ \\
 & & & & 2145 (+2)  & 1.1362  & $-$ \\
 & & & & 2150 (+7)  & 1.1361  & $-$ \\
 & & & & 2162 (+19) & 1.1347  & 2.290 \\
\hline
\end{tabular}
\end{table*}

%
%
\subsection{CO/H$_2$O ice interface: multiple CO adsorption}\label{sec:inter}
\begin{figure*}
\centering
\includegraphics[width=17.0cm]{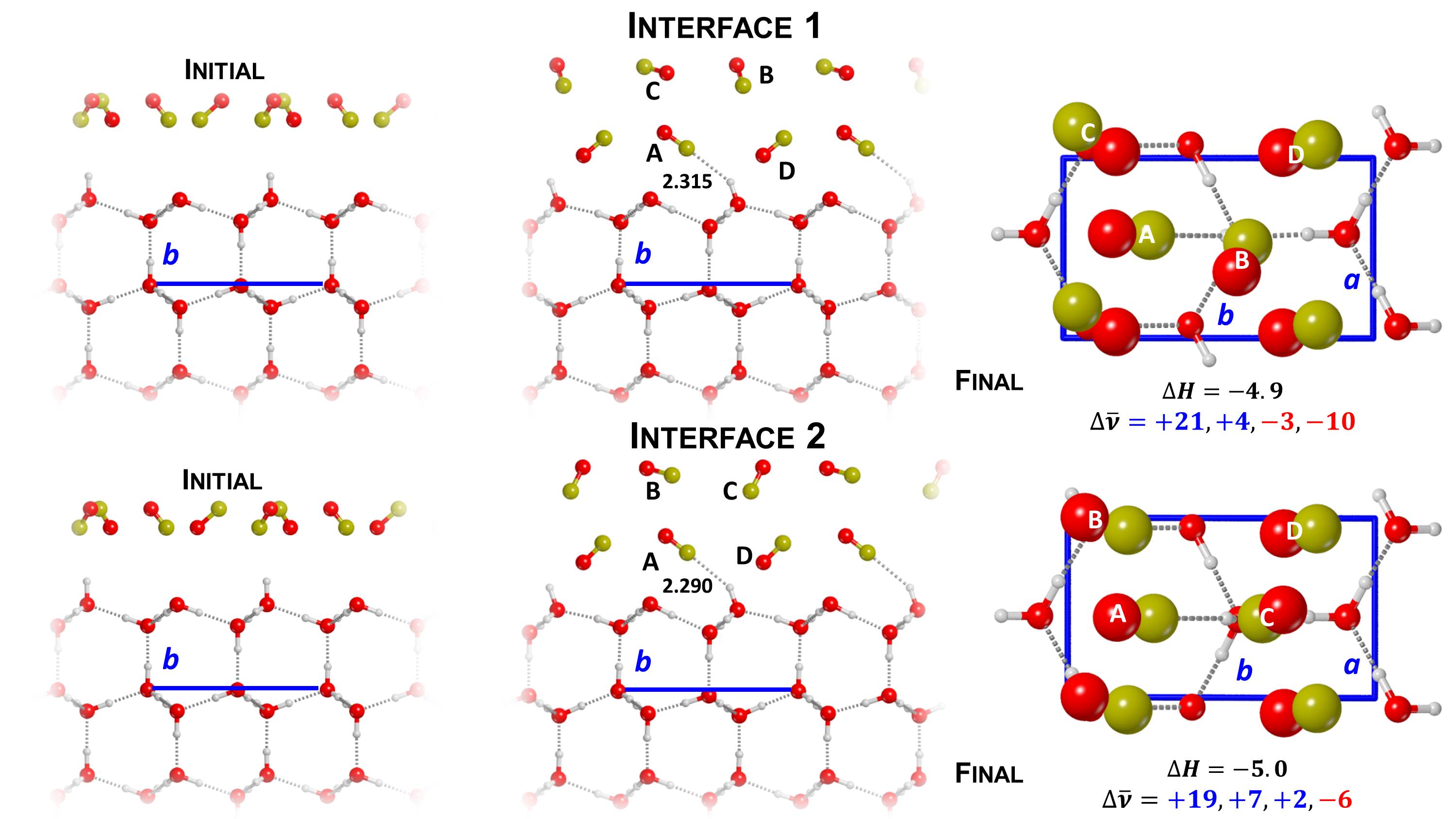}
\caption{CO/H$_2$O interface models, using the (001) surface of \textsc{P-ice}. Left panels: side views along the $\boldsymbol{a}$ direction of the initial geometries. Centre and right panels: final optimized geometries (side views along $\boldsymbol{a}$ direction $-$ centre $-$ and top views $-$ right). In top views, C and O atoms of CO molecules as van der Waals spheres. CO molecules labelled from A to D according to the decreasing order of the vibrational wavenumbers. Adsorption enthalpies {\it per} CO $\Delta H$ in kJ mol$^{-1}$, vibrational shifts $\Delta\bar{\nu}$ in cm$^{-1}$, H-bonds in \AA.}
\label{fig:interface}
\end{figure*}

As the concentration of CO increases up to the monolayer limit, a different kind of model is needed to reproduce the CO/H$_2$O ice interaction. Then, we have designed two different structures by placing four CO molecules {\it per} unit cell upon the (001) surface of \textsc{P-ice}, as in the {\it initial} panels of Figure \ref{fig:interface}.\\
The optimization procedure yields an interesting rearrangement in both CO/H$_2$O ice interfaces. As reported in the {\it final} panels of Figure \ref{fig:interface}, CO molecules reorganize in order to maximize the reciprocal quadrupolar interactions and the H-bonds with the surface. The resulting structures are two CO-layered slabs, resembling the arrangement of CO molecules in the bulk structure. Energetics, computed according to Equations \eqref{eq_DE1} and \eqref{eq_DH}, and spectroscopic features are reported in Table \ref{tab:CO_PICE}.\\
As a consequence of the different surrounding of each CO molecule, the CO stretching band splits in few components and the overall features of the computed spectra are the same for the two interfaces. In particular, adopting the CO labels of Figure \ref{fig:interface}, it can be noted that the high wavenumber peaks, around 2160 cm$^{-1}$, correspond to the interaction of the carbon atom of the CO$_\textsc{A}$ molecule with the \textsc{dH} of the water slabs, as in the case of the \textsc{dH}$\cdot\cdot$CO interaction discussed in Section \ref{sec:onpice}. Similarly to \textsc{dH}$\cdot\cdot$OC, the most redshifted feature is due to the stretching of the CO$_\textsc{D}$ molecules, which are those exposing their O atoms to the H$_2$O ice surface. The two remaining vibrational modes, which involve the CO molecules of the upper layer, are evidently less affected by the icy water surface and reflect the CO-CO interactions.\\
Therefore, it can be stated that, as the availability of \textsc{dH} sites is over, the exceeding CO molecules reorganize and grow up as homogeneous layers on top of the ice surface. As a consequence, a broadening around the 2138 cm$^{-1}$ IR signal takes place due to the different interactions of the adsorbed CO molecules, which leave red- and blueshifts.\\
In very recent works, the \openvirg spontelectric\closedvirg nature of CO layers grown on water ices has been revealed. It has been suggested that this peculiar behaviour could strongly affects the physico-chemical evolution of clouds since it could influence the abundances of charged chemical species in the gas-phase ISM \citep{loto1, loto2}. We hope that our interface models could be useful to computationally highlight this important aspect in further studies.   
%
%
\subsection{CO in H$_2$O ice-cages}\label{sec:cages}
\begin{figure*}
\centering
\includegraphics[width=16cm]{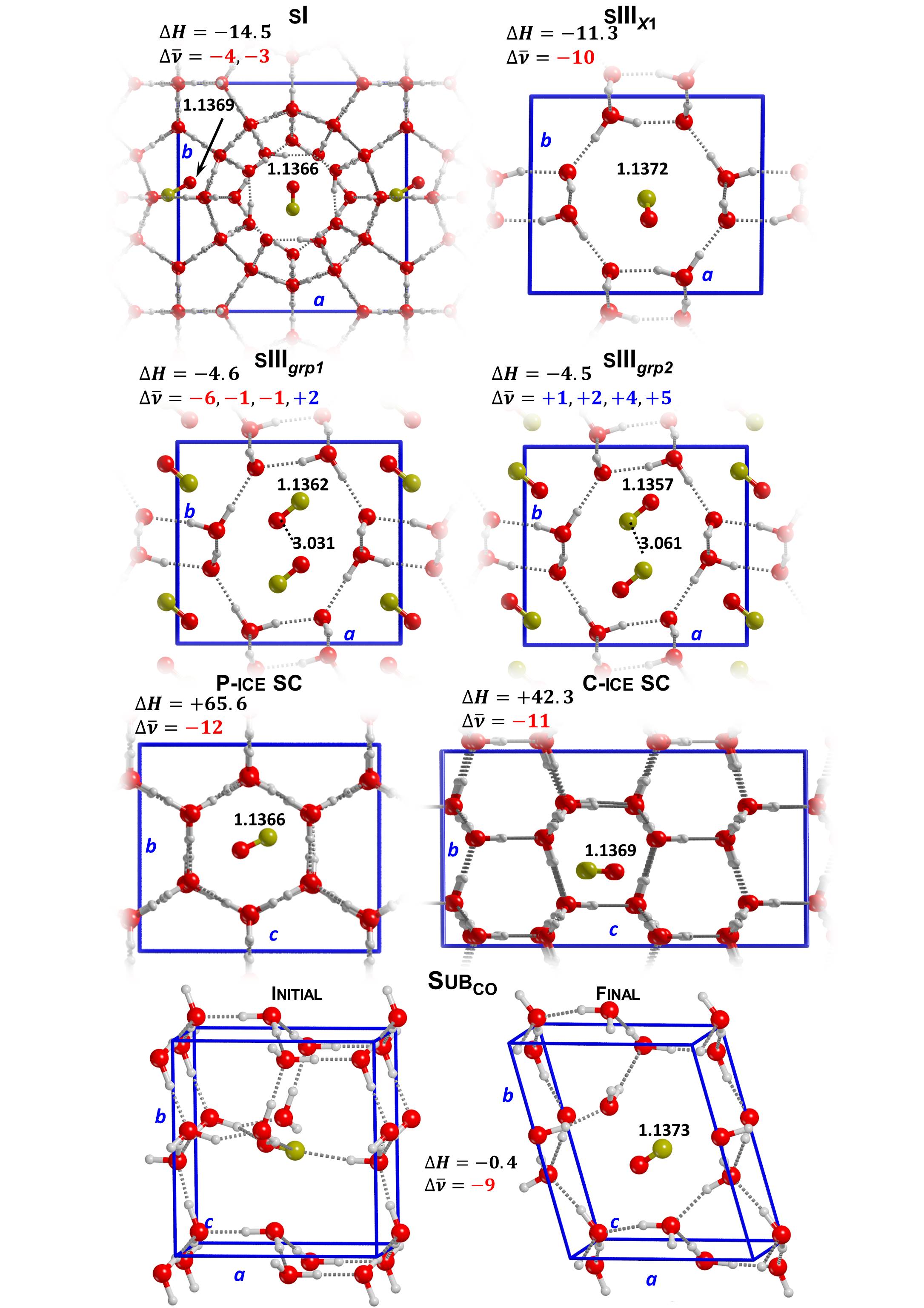}
\caption{CO entrapped in different cage models. Enthalpy variations {\it per} CO molecule $\Delta H$ in kJ mol$^{-1}$, vibrational shifts $\Delta \bar{\nu}$ in cm$^{-1}$, distances in \AA.}
\label{fig:cages}
\end{figure*}

Sandford et al. \citep{AstroJ_Sandford_1988} observed that as the amorphous H$_2$O ice is warmed up, from 10 to 150 K, the H$_2$O molecules rearrange into a more ordered structure. 
Moreover, the annealing process seems, on the one hand, to facilitate CO diffusion into ice pores \citep{JPhysChemA_Palumbo_1997}, as attested by the narrowing of the signal at 2136 cm$^{-1}$ and, on the other hand, it is claimed to be responsible for the appearance of new adsorption features around 2144 cm$^{-1}$ (4.664 $\mu$m) that could be possibly related to the formation of clathrate-like structures \citep{AstroJ_Sandford_1988}.\\
In order to investigate on these aspects, we simulated the CO encagement into different H$_2$O-crystalline lattices ranging from ordinary ice to some clathrates of increasing cavity volume.\\
In particular, with reference to Figure \ref{fig:cages}, we addressed the natural occurring type I CO-clathrate \citep{Clathrate_book, Nature_Davidson_1987}, \textsc{sI}, whose two cages were filled with one CO molecules {\it per} cage, and the recently discovered structure III methane-clathrate \citep{Nature_Loveday_2001}, \textsc{sIII} which, despite its similarity with hexagonal ice, shows larger cavities.
Here we inserted two CO molecules {\it per} cage defining eight different initial configurations and exploiting symmetry. According to their spectroscopic features, these eight initial configurations can be grouped into two groups, labelled as sIII$_{grp1}$ and sIII$_{grp2}$. We also considered a single CO occupancy, \textsc{sIII$_{x1}$}. Finally, the eventuality of CO encagement into crystalline ice was simulated by a 2x2x2 supercell models (SC) of both \textsc{C-} and \textsc{P-ice}. All the systems were fully optimized and the vibrational properties for the CO molecules are summarized in Table \ref{tab:cages}.\\
\begin{table*}
\caption{Energetic, spectroscopic and structural data for CO entrapped in different water-ice cages. Interaction energies $\Delta E^{\textsc{CP}}$ and enthalpies $\Delta H$ {\it per} CO molecule in kJ mol$^{-1}$ ($\Delta E^{\textsc{CP}}$ without the dispersion contribution in parenthesis), wavenumbers $\bar{\nu}$ and shifts $\Delta \bar{\nu}$ with respect to the gas-phase CO stretching in cm$^{-1}$, C$-$O bond lengths $d_{\textsc{C-O}}$ in \AA. \%$\Delta V$ is the percentage variation in cell volume with respect to the initial structure.
\textsc{sIII} cases can be divided in two different groups according to their spectroscopic features (see text). \openvirg Sub$_{\textsc{CO}}$\closedvirg stands for substitutional CO (see text and Figure \ref{fig:cages}). \textsc{sIII}$_{x1}$ refers to only one CO molecule {\it per} cage in \textsc{sIII} structure.}
\label{tab:cages}
\begin{tabular}{lrrrrr}
\hline
Model\,\,\,\,\,\,\,\,\,\,\,\,\,\,\,\,\,\,\,\ &$\Delta E^{\textsc{CP}}$ (no disp.) & \,\,\,\,\,\,\,\,\,\,\,\,\ $\Delta H$ & \,\,\,\,\,\,\,\,\,\,\,\,\ \%$\Delta V$ & \,\,\,\,\,\,\,\,\,\,\,\,\ $\bar{\nu}$ ($\Delta \bar{\nu}$) & \,\,\,\,\,\,\,\,\,\,\,\,\  $d_{\textsc{C-O}}$ \\
\hline
\textsc{sI} 		  			  & $-$16.1 (+4.8) & $-$14.5 & $-$ & 2139 ($-$4) & 1.1369\\
 & & & & 2140 ($-$3) & 1.1366 \\
\textsc{sIII}$_{x1}$   & $-$13.6 (+9.0) & $-$11.3 & +1 & 2133 ($-$10) & 1.1372 \\
\textsc{sIII}$_{grp1}$ & $-$7.7 (+18.2) &  $-$4.6 & +10 & 2137 ($-$6) & $^{\#}$1.1362 \\
 & & & & 2142 ($-$1) & \\
 & & & & 2142 ($-$1)  & \\
 & & & & 2145 ($+$2)  & \\
\textsc{sIII}$_{grp2}$ & $-$7.6 (+17.8) &  $-$4.5 & +10 &  2144 (+1) & $^{\#}$1.1357 \\
 & & & & 2145 ($+$2) &  \\
 & & & & 2147 ($+$4)  & \\
 & & & & 2148 ($+$5) & \\
\textsc{C-ice} SC               & +36.6 (+63.9)  & +42.3   & +2  & 2132 ($-$11) & 1.1366  \\
\textsc{P-ice} SC               & +60.5 (+83.1)  & +65.6   & +3  & 2131($-$12) & 1.1369 \\
\hline
Expulsion 1 & $-$10.9 (+3.1)  & $-$9.7   & & 2140 ($-$3) & 1.1368  \\
Expulsion 2 & $-$7.7 (+13.5)  & $-$5.4   & & 2139 ($-$4) & 1.1367  \\
\hline
Sub$_{\textsc{CO}}$ & $-$3.8 (+18.8) &  $-$0.4 & $-$7 & 2134 ($-$9) & 1.1373 \\
\hline
\end{tabular}
\\
$^{\#}$: all CO molecules have the same $d_{\textsc{C-O}}$ distance. 
\end{table*}

\begin{figure*}
\centering
\includegraphics[width=17cm]{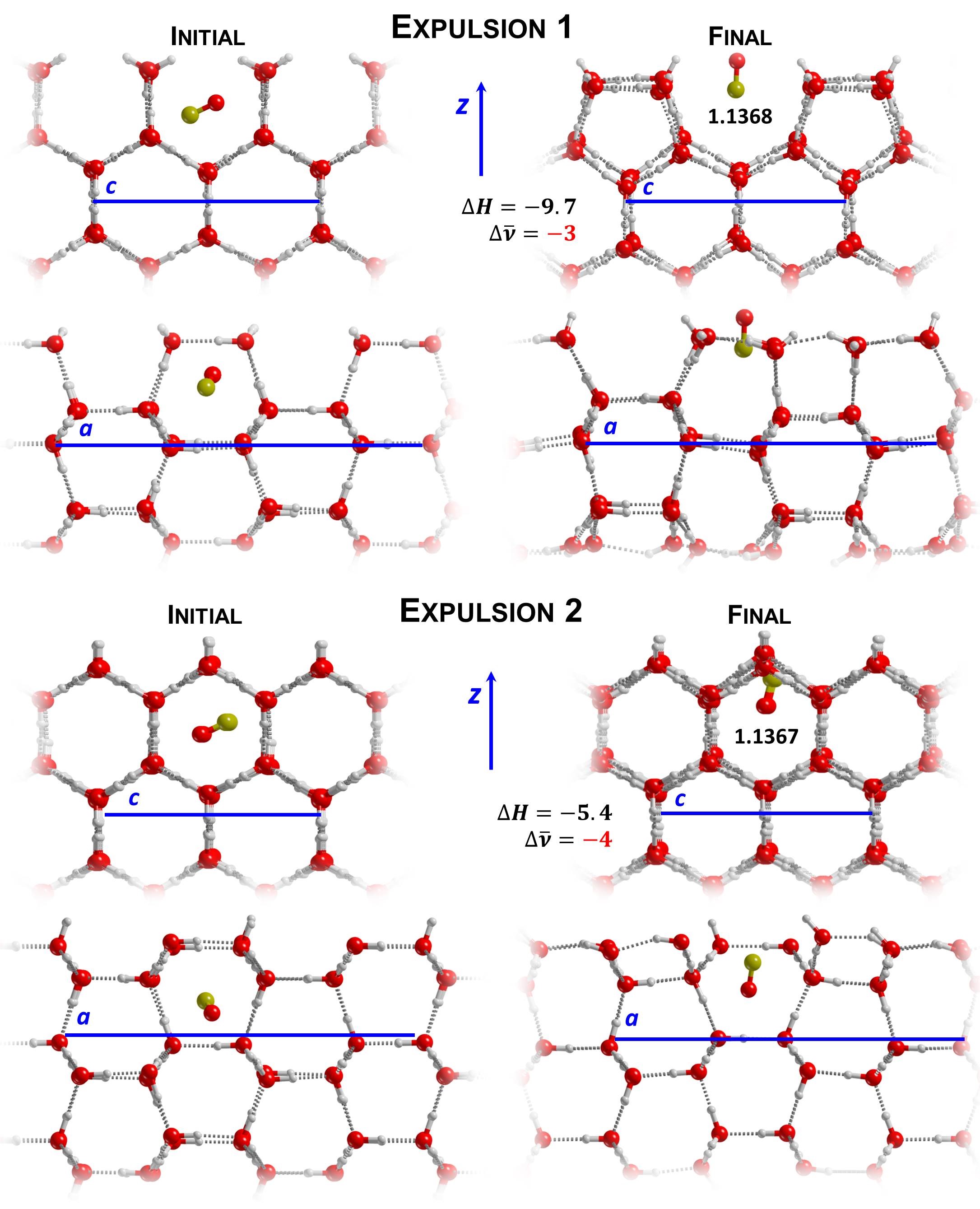}
\caption{Side views for the \openvirg expulsion\closedvirg models, {\it i.e.} CO expelled from superficial \textsc{P-ice} pores. Initial geometries on the left, final ones on the right. Enthalpy variations {\it per} CO molecule $\Delta H$ in kJ mol$^{-1}$, vibrational shifts $\Delta \bar{\nu}$ in cm$^{-1}$, C$-$O distances in \AA.}
\label{fig:expulsion}
\end{figure*}

From our results, some general considerations arise. Data reported in Table \ref{tab:cages} underline the fundamental role of dispersive interactions in determining the energetics of the encagement process. Indeed, there is no evidence of H-bonds involving the C atom of CO molecules with any water molecule of the surrounding so that without the dispersion contribution CO will be unbound. Nevertheless, in the cases of \textsc{C-} and \textsc{P-ice} SC steric hyndrance predominates also when accounting for dispersive forces, {\it i.e.} the 12-H$_2$O molecules cavity is too small to host carbon monoxide, as indicated by the positive interaction energies. Secondly, COs induce a small but sensitive relaxation of the water lattice, resulting in an increase of the cell volume between 1\% to 10\% with respect to the empty initial structures in all cases but \textsc{sI}.\\
The analysis of spectroscopic data shows that the formation of different clathrate-like cages around CO molecules can partially explain the observed IR signals for CO in water-dominated environments. The redshift of the CO stretching wavenumber with respect to the gas-phase is significantly less than in the case of \textsc{dH}$\cdot\cdot$OC. This behaviour can again be explained by invoking quadrupolar interactions between the ice lattice and the CO. The extent of the redshifting effect for those cases where a single cage hosts a single CO molecule, ({\it i.e.} \textsc{sI}, \textsc{sIII}$_{x1}$, Sub$_\textsc{CO}$, \textsc{C-} and \textsc{P-ice} SC) seems to suggest a general trend between volume of the cage and the redshift itself. However, the volume of the cage cannot be properly defined and thus we performed a detailed neighbouring analysis for the C and O atom of COs. All data are reported in the \openvirg Neighbourhood analysis for CO in water cages\closedvirg section of the Appendix A (Table A6). Briefly, results indicate that those cases characterized by very small cages (as \textsc{P-} and \textsc{C-ice} SCs) have very positive interaction energies and present the most redshifted wavenumbers, while when the dimensions of the cages are larger (as in the \textsc{sI} case) the entity of the redshift is small.\\
A noticeable difference appears in the eight \textsc{sIII} configurations explored. Despite their similarity in terms of CO-CO and CO-lattice distances, see Figure \ref{fig:cages}, in the \textsc{sIII}$_{grp2}$ all the four CO stretching modes are blueshifted, while in the \textsc{sIII}$_{grp1}$, only one mode is blueshifted.\\
To enlighten on this aspect, we removed the symmetry and reoptimized the \textsc{sIII}$_{grp1}$ and \textsc{sIII}$_{grp2}$ structures in $P1$ space group, ending up with the same frequency pattern. For a more detailed analysis of the IR signals with and without symmetry, see Table A5 of the Appendix A. We conclude that these blueshifted modes are not a symmetry artefact but the result of the peculiar geometry assumed by the CO molecules inside the cages giving rise to specific intermolecular interactions. As a matter of fact, these signals are completely absent in the single occupied \textsc{sIII$_{x1}$} system.\\
Our data suggest that both the astronomical features of CO/H$_2$O mixtures and the controversial redshifted peaks in the experimental spectra could be due to trapped CO experiencing the potential field inside a water cage and/or {\it via} quadrupolar interactions of neighbouring CO molecules.\\
On the other hand, the very high energetic data for CO entrapped in crystalline \textsc{C-} and \textsc{P-ice} suggest that carbon monoxide could not be able to diffuse into such small pores but does prefer to remain adsorbed onto proper surface moieties or eventually to rearrange the network of water molecules around it.\\
In order to test this possibility, from the 3D periodic structure of the \textsc{P-ice} SC of Figure \ref{fig:cages}, we cut two different slab models along the [010] direction so that the water cages hosting COs result to be exposed to vacuum. Surfaces are more flexible structures than bulks so they represent natural escaping routes. The initial and final geometries of these two models are shown in Figure \ref{fig:expulsion}, while the computed properties are reported  in Table \ref{tab:cages}, \openvirg Expulsion\closedvirg rows. Once again, the role of dispersive interactions is fundamental to determine the final energetics and structural features. CO molecules tend to be expelled from the surface pores and only weak quadrupole interactions keep them inside the structures. However, these \openvirg expulsions\closedvirg cause a great deformation of the surrounding lattice and neighbour water molecules rearrange into \openvirg semi-cages\closedvirg partially restoring clathrate-like cavities.\\
These simulations suggest an \openvirg hydrophobic behaviour\closedvirg of CO and clearly show that CO would hardly diffuse into water-ice pores unless they are at least as large as in clathrates. This behaviour is confirmed by the \openvirg substitutional CO\closedvirg case (Sub$_{\textsc{CO}}$ in Figure \ref{fig:cages} and Table \ref{tab:cages}), where one water molecule of the \textsc{P-ice} bulk structures is replaced with one CO interacting, {\it via} H-bonds, with two H$_2$O molecules. Once the structure is relaxed, water lattice  re-arrange significantly, CO$\cdot\cdot$H$_2$O directional bonds are lost and a clathrate-like cage forms around the carbon monoxide (compare {\it initial} and {\it final} Sub$_{\textsc{CO}}$ panels of Figure \ref{fig:cages}). The energy balance, reported in the last row of Table \ref{tab:cages}, takes into account both the deformation energy and the formation of new H-bonds between H$_2$O molecules and results in an almost zero value. This result suggests that the formation of a CO/H$_2$O mixed lattice is rather unlikely, at least at low concentration of carbon monoxide and at the very low temperature characterizing the colder regions of the ISM.\\
%
%
\subsection{CO entrapped in water-rich mixture}\label{sec:dirty}
Sub$_{\textsc{CO}}$ can be considered as the simplest model of a CO/H$_2$O mixed ice. In order to improve the description of these binary systems, we defined a more realistic model by replacing 6 out of 24 water molecules by 6 CO molecules in a 1x1x2 supercell of \textsc{P-ice}  to reproduce the CO:H$_2$O ratio usually reported for the ISM \citep{ARAA_Boogert_2015}.
The initial geometry was amorphized by running an AIMD simulation with the CP2K code at 300 K for 15.0 ps. After that, 5 snapshots were selected, corresponding to 5.0, 7.5, 10.0, 12.5 and 15.0 ps of production and each structure was optimized at PBE-D2 level with the CRYSTAL17 code. On the final optimized geometries, a complete energy and frequency analysis was performed. Data reported in Table \ref{tab:mixed}, confirm once again the fundamental role of dispersive  interactions to define the energetic configuration of these systems where, as in the case of Sub$_{\textsc{CO}}$, clathrate-like structures build up around CO molecules.\\
\begin{figure*}
\centering
\includegraphics[width=17cm]{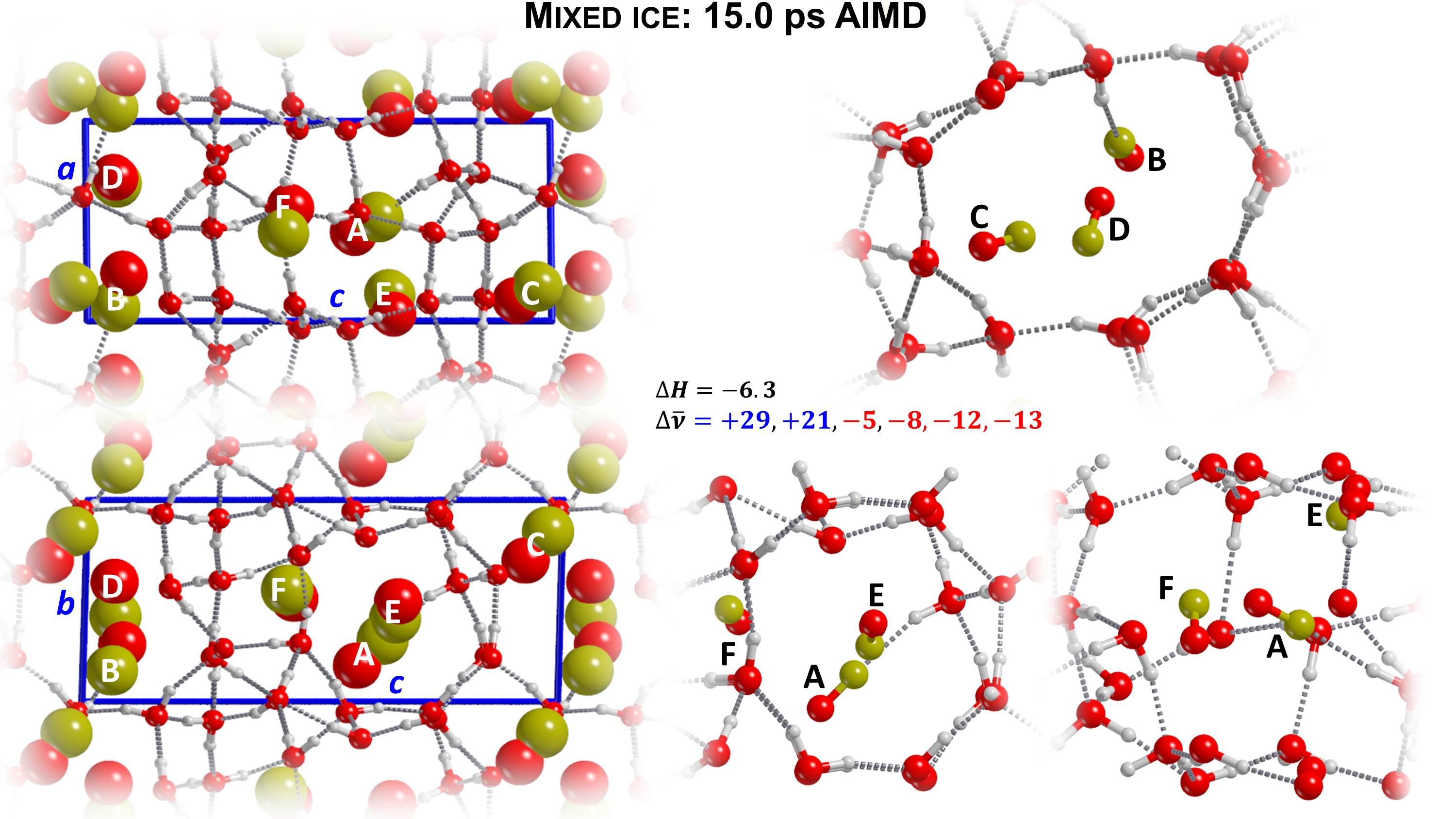}
\caption{Left: view along the $\boldsymbol{b}$ (top) and $\boldsymbol{a}$ (bottom) lattice vector for the mixed CO/H$_2$O system. C and O atoms of CO molecules as van der Waals spheres. 
Right: detailed views of the local water-environments surrounding CO molecules. CO molecules labelled from A to F according to their stretching wavenumbers. Enthalpy variation {\it per} CO molecule $\Delta H$ in kJ mol$^{-1}$, vibrational shifts $\Delta \bar{\nu}$ in cm$^{-1}$.} 
\label{fig:15ps}
\end{figure*}
\begin{figure*}
\centering
\includegraphics[width=17.5cm]{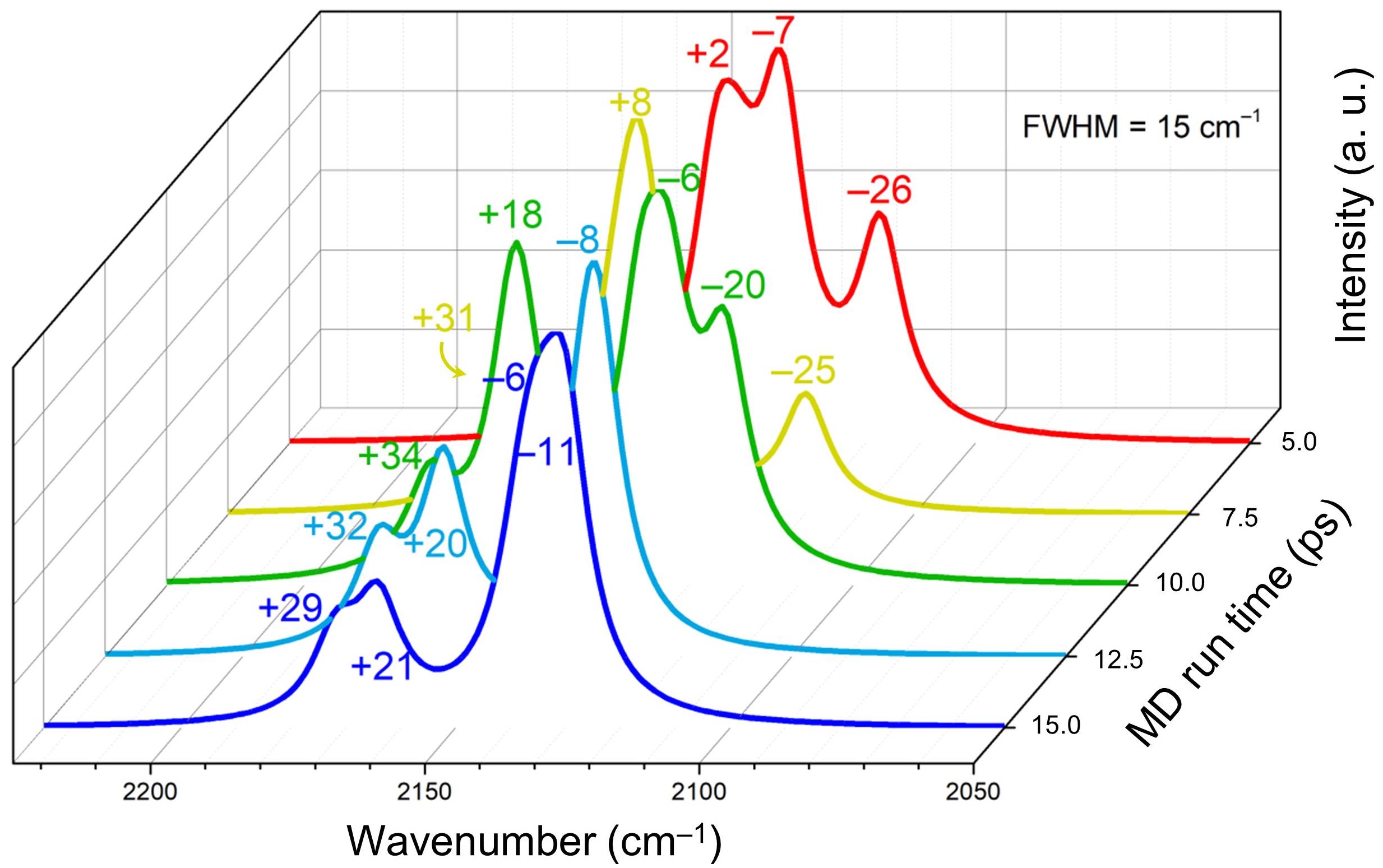}
\caption{Simulated IR spectra for the different AIMD snapshots as a function of the AIMD production time. The values of the vibrational shifts (in cm$^{-1}$) are reported close to all distinguishable peaks. The FWHM for each component of the different spectra is set equal to 15 cm$^{-1}$.}
\label{fig:spectra}
\end{figure*}
The 15.0 ps optimized structure, see Figure \ref{fig:15ps}, turns out to be a reliable model of a possible structure since its spectrum closely resembles the experimental ones; therefore, we performed a detailed energetic analysis to determine the interaction energies and enthalpies of each individual CO molecule within the mixed CO-ice lattice, Table \ref{tab:mixed}. These quantities were estimated according to Equations \eqref{eq_DE1} and \eqref{eq_DH} by setting to zero the deformation energy of the mixed [18 H$_2$O $+$ 5 CO] system.\\
As in the Sub$_{\textsc{CO}}$ case, the water lattice tends to form cage structures close resembling the natural cavities in clathrates. Thus, CO molecules entrapped within these cages, and not involved in H-bonds, are characterized by redshifts. However, during the evolution of the simulation, H-bonds among CO and water can actually form in our model. The CO molecules involved in these H-bonds have, in accordance with our other results on \textsc{P-ice} slabs, quite large blueshifts. The absolute computed IR intensities for this blueshifted signals are smaller than those for redshifted ones. The lack of the 2152 cm$^{-1}$ peak $-$ that together with our other previous results was assigned to the \textsc{dH}$\cdot\cdot$CO interaction $-$ might be explained by a low concentration of \textsc{dH} sites in water-dominated ices grown under interstellar conditions.\\
Referring to Figure \ref{fig:15ps} and using the adopted notation, it can be seen that the A and B CO molecules, involved in H-bonds, have the highest strength of interaction whereas for all the other COs the energetic is consistent with dispersion-driven interactions (Table \ref{tab:mixed}). It is worth noting that the surroundings of A and B CO molecules are very different from the \textsc{dH}$\cdot\cdot$CO ones as occurring on \textsc{P-ice} surfaces since these molecules are entrapped in water cages.\\
To appreciate the evolution of the spectroscopic signals, the five IR spectra of the corresponding AIMD snapshots are superimposed in Figure \ref{fig:spectra}. All spectra except the 7.5 ps one present a main persistent component at $\approx -$7 cm$^{-1}$. A more redshifted feature at $\approx -$20 cm$^{-1}$ rapidly disappears, while after 7.5 ps, two blueshifted signals appear at around +20 and +30 cm$^{-1}$, but while the former is preserved in the final spectrum, the latter also disappears, resulting in a shoulder of the main blueshifted peak.\\
\begin{table*}
\caption{Structural, energetic and spectroscopic features for the five snapshots selected from the AIMD run for the mixed CO/H$_2$O ice system. AIMD times in ps, energetic quantities {\it per} CO in kJ mol$^{-1}$, wavenumbers and shifts in cm$^{-1}$, distances in \AA. For the 15.0 ps, we computed the energetic features for all the six CO molecules, individually (see text).} 
\label{tab:mixed}
\begin{tabular}{lcrrrrr}
\hline
\multicolumn{2}{l}{MD time}  & \,\,\,\,\,\, $\Delta E^{\textsc{CP}}$ (no disp.) & \,\,\,\,\,\,\,\,\,\,\,\, $\Delta H$ & \,\,\,\,\,\,\,\,\,\,\,\, $\bar{\nu}$ ($\Delta \bar{\nu}$) & \,\,\,\,\,\,\,\,\,\,\,\, $d_{\textsc{CO}}$ & \,\,\,\,\,\,\,\,\,\,\,\, H-bond \\
\hline
\multicolumn{2}{l}{5.0}  & $-$2.2 (+13.7)& +0.5     & 2117 ($-$26)$\div$2148 (+5)   & 1.1392$\div$1.1359 & 2.021, 2.394 \\
\multicolumn{2}{l}{7.5}  & $-$10.6 (+3.4)& $-$7.7   & 2120 ($-$23)$\div$2174 (+31)  & 1.1390$\div$1.1330  & 2.006, 2.207, 2.460  \\
\multicolumn{2}{l}{10.0} & $-$6.5 (+9.9) & $-$3.4   & 2123 ($-$20)$\div$2177 (+34)  & 1.1386$\div$1.1325  & 1.937, 2.150 \\
\multicolumn{2}{l}{12.5} & $-$7.3 (+7.2) & $-$7.1   & 2135 ($-$8)$\div$2175 (+32)   & 1.1371$\div$1.1329  & 2.013, 2.078  \\
\multicolumn{2}{l}{15.0} & $-$9.5 (+8.6) & $-$6.3   &  &  &     \\
          & A     & $-$25.9 (0.0)    & $-$21.7  & 2172 (+29)    &  1.1333  & 2.036  \\
          & B     & $-$23.9 ($-$2.8) & $-$20.3  & 2164 (+21)    &  1.1341  & 2.055  \\
          & C     & $-$12.8 (+11.1)  & $-$10.1  & 2138 ($-$5)   &  1.1369  & $-$    \\
          & D     & $-$15.5 (+6.8)   & $-$12.9  & 2135 ($-$8)   &  1.1374  & $-$    \\
          & E     & $-$18.8 (+6.9)   & $-$15.8  & 2131 ($-$12)  &  1.1376  & $-$    \\
          & F     & $-$5.6 (+23.8)   & $-$1.8   & 2130 ($-$13)  &  1.1374  & $-$    \\
\hline
\end{tabular}
\end{table*}

%
%

\subsection{Merging the IR signals: global simulated spectrum}
\begin{table*}
\centering
\caption{Resume of our energetic and spectroscopic computed data. Enthalpy variations {\it per} CO $\Delta H$ in kJ mol$^{-1}$, IR stretching wavenumbers $\bar{\nu}$ and shifts $\Delta \bar{\nu}$ in cm$^{-1}$. Absolute IR intensities (in km mol$^{-1}$) {\it per} CO are also reported.}
\label{tab:resume_comp_res}
\begin{tabular}{lrrrrlrrrr}
\hline
Model & Figure & $\Delta H$ & $\bar{\nu}$ ($\Delta \bar{\nu}$) & Intensity & Model & Figure & $\Delta H$ & $\bar{\nu}$ ($\Delta \bar{\nu}$) & Intensity\\
\hline
Gas phase CO & \ref{fig:pureco} & $-$ & 2143 (0) & 62.04 & sI & \ref{fig:cages} & $-$14.5 & 2139 ($-$4) & 23.18\\     
HOH$\cdot\cdot$CO & A20  & $-$ & 2159 (+16)  & 60.86 &  &  &  & 2140 ($-$3) & 181.04\\  
HOH$\cdot\cdot$OC & A20  & $-$ & 2136 ($-$7) & 78.28 & sIII$_{grp1}$ sym. & \ref{fig:cages} & $-$4.6 & 2137 ($-$6) & 78.36\\  
Solid $\alpha$-CO symm. & \ref{fig:pureco} & $^{\#}-$8.0 & *2139 ($-$4) & 0.00 &  &  & & 2142 ($-$1) & 49.88\\  
  &  &  & 2142 ($-$1) & 94.00 &  &  &  & 2142 ($-$1) & 0.00\\  
Solid $\alpha$-CO no symm. & \ref{fig:pureco} & $^{\#}-$9.0 & 2135 ($-$8) & 110.83 &  & &  & *2145 (+2)& 0.00\\  
  &  &  & 2136 ($-$7) & 109.40 & sIII$_{grp2}$ symm. & \ref{fig:cages} & $-$4.5 & 2144 (+1) & 55.94\\  
  &  &  & 2144 (+1) & 0.25 &  &  & & 2145 (+2) & 0.00\\  
  &  &  & 2145 (+2) & 0.03 &  &  & & 2147 (+4) & 44.62\\  
\textsc{P-ice}: (010) \textsc{dH}$\cdot\cdot$CO & \ref{fig:010} & $-$18.4 & 2158 (+15) & 52.49 & &  & & *2148 (+5) & 0.00\\  
\textsc{P-ice}: (010) \textsc{dH}$\cdot\cdot$OC & \ref{fig:010} & $-$12.9 & 2114 ($-$29) & 92.98 & \textsc{C-ice} SC& \ref{fig:cages} & +42.3 & 2132 ($-$11) & 135.72\\  
\textsc{P-ice}: (001) \textsc{dH}$\cdot\cdot$CO & \ref{fig:001} & $-$16.8 & 2163 (+20) & 49.07 & \textsc{P-ice} SC & \ref{fig:cages} & +65.6 & 2131 ($-$12) & 139.15\\  
\textsc{P-ice}: (001) \textsc{dH}$\cdot\cdot$OC & \ref{fig:001} & $-$13.3 & 2122 ($-$21) & 90.15 & Sub$_{\textsc{CO}}$ & \ref{fig:cages} & $-$0.4 & 2134 ($-$9) & 143.96\\  
\textsc{P-ice}: interface 1 & \ref{fig:interface} & $-$4.9 & 2133 ($-$10) & 106.62 & \textsc{P-ice}: expulsion 1 & \ref{fig:expulsion} & $-$9.7 & 2140 ($-$3) & 100.2\\  
  &  &  & 2140 ($-$3) & 70.54 & \textsc{P-ice}: expulsion 2 & \ref{fig:expulsion} & $-$5.4 & 2139 ($-$4)& 85.59\\  
  &  &  & 2147 (+4) & 49.41 & Mixed ice: 15.0 ps AIMD & \ref{fig:15ps} & $-$6.3 & 2130 ($-$13) & 179.18\\  
  &  &  & 2164 (+21) & 48.88 &  &  & & 2131 ($-$12) & 66.05\\  
\textsc{P-ice}: interface 2 & \ref{fig:interface} & $-$5.0 & 2137 ($-$6) & 87.07 & & &  & 2135 ($-$8) & 127.69\\  
  &  &  & 2145 (+2) & 81.81 & & & & 2138 ($-$5) & 101.14\\  
  &  &  & 2150 (+7) & 33.69 & & & & 2164 (+21) & 97.14\\  
  &  &  & 2162 (+19) & 46.43 & & & & 2172 (+29) & 67.92\\  
sIII$_{x1}$ & \ref{fig:cages} & $-$11.3 & 2133 ($-$10) & 95.82 & & & & & \\  
\hline    
\end{tabular}
\\
$^{\#}$: sublimation enthalpy.\\
*: IR inactive mode.
\end{table*}

\begin{figure*}
\centering
\includegraphics[width=17cm]{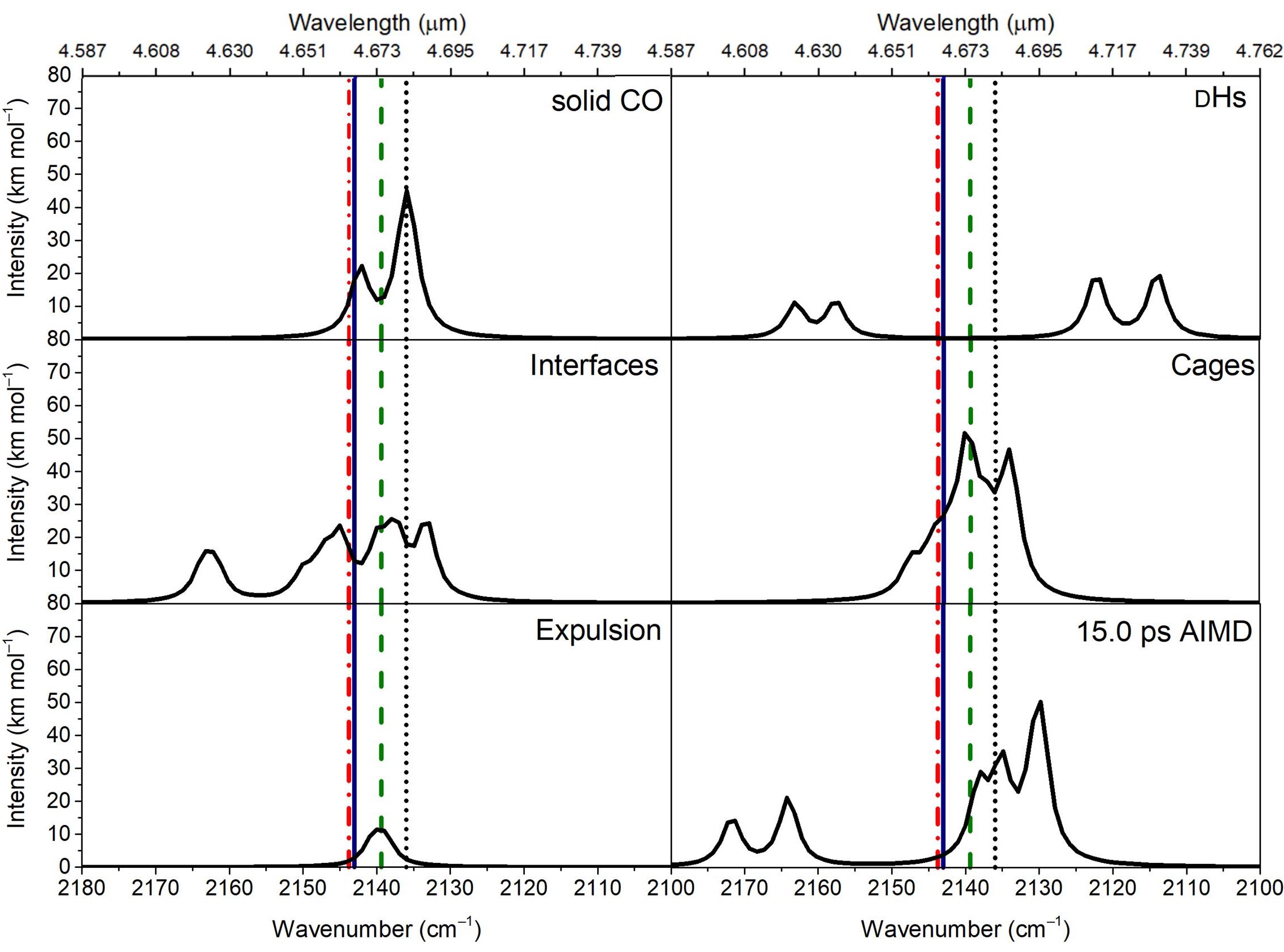}
\caption{Overall simulated spectra for each category. The FWHM of each component is set equal to 3 cm$^{-1}$. In the \openvirg cages\closedvirg panel, we excluded the SC contributions because of energetic reasons. The straight lines represent: gas-phase CO stretching at 2143 cm$^{-1}$ (blue continuous line) and the \textsc{Ap} and \textsc{Pp} components reported by \citet{ARAA_Boogert_2015} at 2143.7, 2139.9 and 2136.5 cm$^{-1}$ (red dot-dashed, green dashed and black dotted lines, respectively).}
\label{fig:spectra_sep}
\end{figure*}

\begin{figure*}
\centering
\includegraphics[width=17cm]{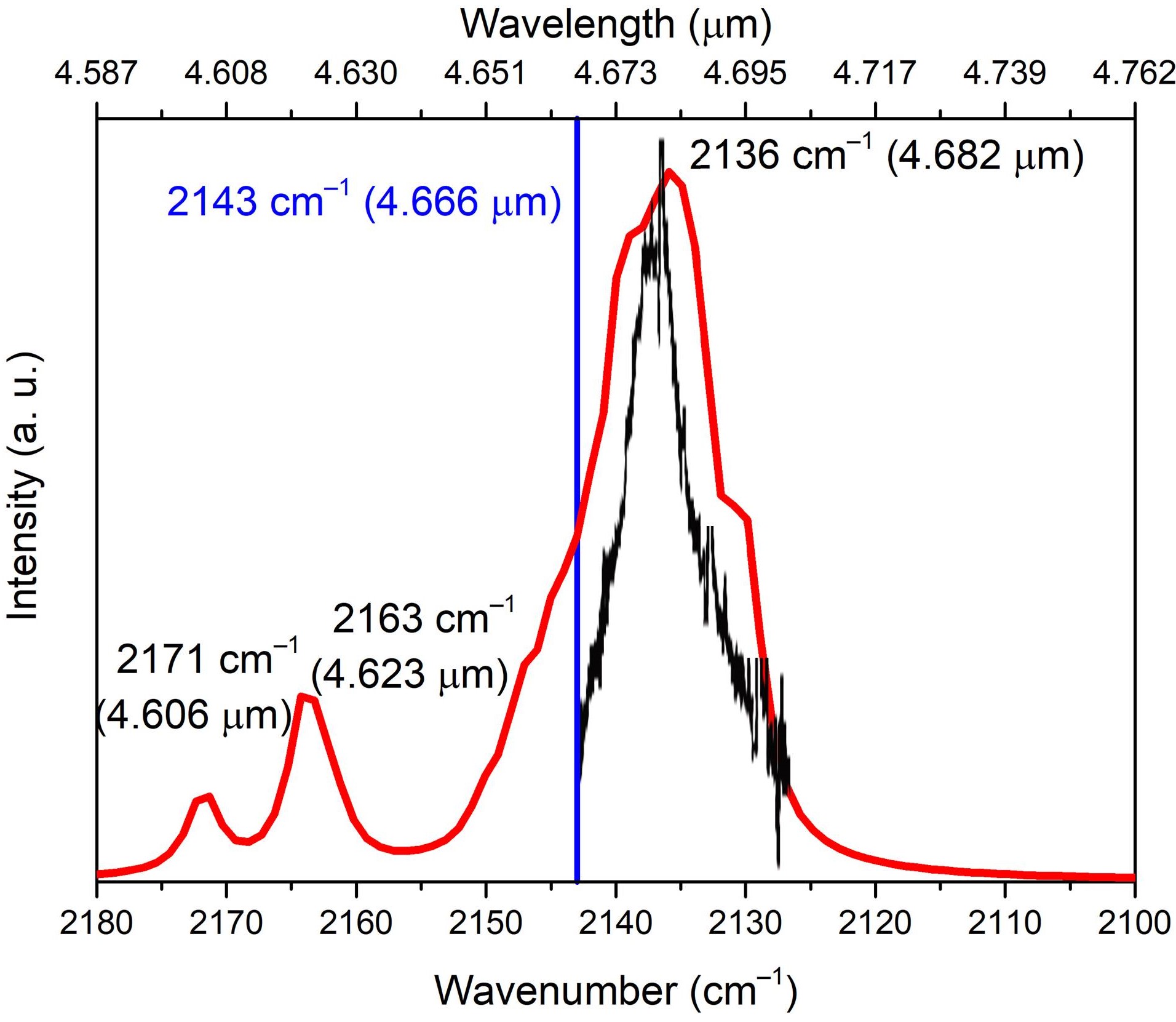}
\caption{Overall simulated spectrum obtained as a sum of all panels of Figure \ref{fig:spectra_sep} expect the \textsc{dHs} one. The FWHM of each component is set equal to 3 cm$^{-1}$. In blue, free gas-phase CO stretching reference. The maximum of the signal is at 2136 cm$^{-1}$ (4.682 $\mu$m). The solid black line represents the observed spectrum for LYSO L 1489 IRS object. This spectrum is taken and adapted from \citet{ARAA_Boogert_2015}.}
\label{fig:spectrum_def}
\end{figure*}

As a general observation, the computed absolute infrared intensities for the CO blueshifted frequencies are lower than those for redshifted ones (see Tables \ref{tab:resume_comp_res} and A4). This is the result of an \openvirg enhancement process\closedvirg which can be underlined by comparing the absolute intensities of a single CO interacting {\it via} its C and O atom with the H atom of a water molecule (Figure A20 of the Appendix A). In the first case (H-bond with the C atom), the CO stretching wavenumber is blueshifted by 16 cm$^{-1}$ and has an absolute intensity of 60.86 km mol$^{-1}$ (see HOH$\cdot\cdot$CO line in Table \ref{tab:resume_comp_res}), slightly lower than that for gas-phase CO (62.04 km mol$^{-1}$). On the other hand, when an H-bond forms between water and the O atom of the CO molecule, the CO stretching wavenumber is redshifted by 7 cm$^{-1}$ and the absolute intensity is enhanced by more than the 25\% (78.28 km mol$^{-1}$) with respect to the gas-phase (HOH$\cdot\cdot$OC line in Table \ref{tab:resume_comp_res}). This in accordance with previous results where a similar enhancement has been observed for CO interacting with selected metal ions \citep{ferrari_CO}. This enhancement of redshifted wavenumbers could partially explain the lack of the blueshifted component in astronomical spectra which is very faint and partially hidden.\\
In Figure \ref{fig:spectra_sep} we reported the overall spectra obtained by summing the computed absolute intensities {\it per} CO molecule ({\it i.e.} the absolute intensity of a specific vibrational mode divided by the number of COs involved in that specific mode) of all tested cases for each category. The FWHMs for each component of the simulated spectra were arbitrarily set equal to 3 cm$^{-1}$; this value is comparable with those reported for the three components of the 4.65 $\mu$m CO band in interstellar ices \citep{ARAA_Boogert_2015}. In the \openvirg cages\closedvirg panel, we {\it a priori} excluded SC models because of their highly unfavourable interaction enthalpies (Tables \ref{tab:cages} and \ref{tab:resume_comp_res}). For a better comparison, we also highlighted the 2143 cm$^{-1}$ (4.666 $\mu$m) free CO signal (blue continuous line) as well as the 2143.7, 2139.9 and 2136.5 cm$^{-1}$ (4.665, 4.673 and 4.681 $\mu$m) ones corresponding to \textsc{Ap} (CO$_2$/CO $>$ 1 and CO $>$ 90\%) and \textsc{Pp} interstellar ices (red dot-dashed, green dashed and black dotted lines, respectively $-$ \citet{ARAA_Boogert_2015}).
However, it is worth underlining the fact that our models cannot be interpreted in terms of \textsc{Ap} and \textsc{Pp} ices, mainly because we did not tested proper mixed ices ({\it i.e.} also including other species such as CO$_2$, CH$_3$OH, CH$_4$ etc.); indeed, our aim was to elucidate the structural, energetic and spectroscopic features of CO in different water ice structural environments. Nevertheless, some conclusions can be drawn as both the \textsc{Apolar} and \textsc{Polar} components can result from CO involved in very different physico-chemical situations, from pure CO systems (solid CO panel) to mixed ones (15.0 ps AIMD panel), passing from layered and encaged models.\\    
We merged all our results in the single, overall spectrum of Figure \ref{fig:spectrum_def}, that is, the sum of all panels of Figure \ref{fig:spectra_sep} excluding the \textsc{dHs} one. \textsc{dH} cases have been excluded because they envisage slab models that underestimate the typical CO:H$_2$O ratio in the ISM \citep{ARAA_Boogert_2015}. For sake of semi-quantitative comparison, the observed spectrum belonging to the LYSO L 1489 IRS object from \citep{ARAA_Boogert_2015} was superimposed. The prediction of the maximum of the peak is in good agreement with the astronomical and experimental observations, even if the band width is slightly overestimated. Blueshifted peaks are visible in the computed spectrum with intensities definitely lower than the main peak. As we have already addressed, the reason for the missing of these blueshifted signals in astronomical spectra is probably due to the low concentration of \textsc{dH} sites in interstellar ices, together with what previously asserted for the IR intensities of the \textsc{dH}$\cdot\cdot$CO interaction. An interesting and alternative explanation to this problem provides that \textsc{dH} sites could be involved in stronger interactions with other molecular species \citep{Fraser_2004}, with particular reference to CO$_2$ \citep{Garrod_Pauly}.\\ 
From a pure energetic point of view, our results underline how CO is only physisorbed to the water ice matrix. Indeed, the enthalpy values fall in a $\sim$ 5-18 kJ mol$^{-1}$ range (excluding SC models and the Sub$_\textsc{CO}$ one where a large part of the interaction energy is consumed in the water lattice rearrangement) and thus weak interactions such as dispersive and electrostatic ones are prominent. Our data can be compared with those usually inferred from classical thermal programmed desorption (TPD) experiments, where different monolayers (MLs) of CO are initially adsorbed onto selected surfaces, such as water ice ones, and then desorbed. Referring to \citet{Smith_2016}, the desorption energies for CO adsorbed on amorphous solid water at very low coverages ($<$ 0.05 ML) are slightly greater than 15 kJ mol$^{-1}$, in good with our computed $\Delta H$ values for \textsc{dH} cases, in particular the \textsc{dH}$\cdot\cdot$CO ones. On the opposite side, the desorption energies for two monolayers can be derived to be approximately equal to 7 kJ mol$^{-1}$, to be compared with our \openvirg \textsc{P-ice} interface\closedvirg models ($\sim$ 5 kJ mol$^{-1}$, Tables \ref{tab:CO_PICE} and \ref{tab:resume_comp_res}). However, please note that a proper comparison cannot be performed because of the lack of a generic definition of monolayer, both from an experimental and computational point of view.\\
All energetic and spectroscopic data are summarized in in Table \ref{tab:resume_comp_res}. These data can be useful to the reader to rebuild the overall spectrum with different values of the FWHM.
%
\section{Resume and conclusions}\label{sec:conclusions}
Adsorption of organic molecules on or inside ice surfaces of the dust mantle of interstellar grains have received significant attention in astrochemical investigations due to their implication in reactions yielding precursor molecules for life. Specifically, from a chemical point of view, the interest around CO is mainly due to the possible formation of derived species such as carbon dioxide, methanol, formaldehyde and formic acid \citep{Karssemeijer_apj_2014}. In the meanwhile, impressive quantity of spectroscopic measurements has become available and several laboratory and theoretical simulations have crossed their intuition to provide reliable molecular configurations to be associated with each spectroscopic peak.
As regards CO, the debate is mainly concentrated on the red- and blueshifts of its stretching frequency recorded as ISM signals with respect to its characteristic gas-phase frequency.\\
Then, in order to enlighten on this very small and informative portion of the spectrum, we simulated a wide range of CO/H$_2$O ice environments adopting different models of water ice ({\it i.e.} crystalline and amorphous ice bulk, surfaces, clathrates) by means of density functional theory simulations (PBE-D2 level of theory) on periodic and molecular models. Our main goal was to furnish solid theoretical justifications on this topic, with particular reference to the interpretation of both the experimental and astronomical spectra. Because the vibrational shifts involved are very small, we ensured that the calculations were run with an accurate level of theory.\\
The present results, far from being conclusive, provide some general indications on CO/H$_2$O ice binary system, both as regards its properties and spectroscopic fingerprint: 
\begin{enumerate}
\item dispersive and quadrupolar interactions have a prominent role in determining the structural and spectroscopic features of carbon monoxide interacting with other systems, {\it e.g.} other CO and/or water molecules. Within our computational methodology, the electrostatic properties of the CO molecule (orientation of the dipole and quadrupole electric moments) are correctly described, even if the value of the CO quadrupole is underestimated with respect the experimental one;
\item the most favourable interaction between carbon monoxide and water involves dangling H atoms (\textsc{dH}) of crystalline  H$_2$O ice (\textsc{P-ice}) surface models and the C atom of CO (negative pole of both dipole and quadrupole). It is this specific  \textsc{dH}$\cdot\cdot$CO interaction that causes a blueshift of the CO stretching wavenumber with respect to the gas-phase value. The symmetric \textsc{dH}$\cdot\cdot$OC interactions is weaker and is lost during the geometry optimizations $-$ as the coupled action of dispersive and quadrupolar forces overcomes the \textsc{dH}H$\cdot\cdot$OC bond, with a corresponding redshift of the C$-$O stretching frequency. The large shift values (in module) for these reference CO/H$_2$O ice models (in particular the large redshifts), together with the very low CO/H$_2$O ratio, suggest that they are probably not representative of a real astronomical ices, at least considering the need of a crystalline ice;
\item the lack of the blueshifted  peak in astronomical spectra could be the result of two factors: firstly, blueshifted signals are intrinsically less intense than redshifted ones and, secondly, \textsc{dH} sites are probably scarcely present (but not completely absent) in the interstellar ices covering dust grain cores, probably due to the amorphous nature of ice in which all available \textsc{dH} are engaged in H-bond to maximize the interaction energy within the ice mantle. Please note that alternative explanations to this problem have been also suggested;
\item the redshifted peak around 2138 cm$^{-1}$ present in both astronomical and experimental spectra could be the result of CO involved in different reciprocal configurations with respect to the water-ice matrix such as layered CO/H$_2$O ices, CO entrapped within water cages of different sizes, pure (crystalline-like) solid CO or proper mixed water ices. For all these models, small redshifts appear as a result of the establishment of weak, non H-bond interactions. However, these interactions could also produce (in all cases but solid CO) small blueshifted signals;
\item CO shows a \openvirg hydrophobic behaviour'', which hinders it to penetrate into small ice pores. As a consequence, the presence of CO can induce large rearrangements of the surrounding lattice water molecules, resulting in the formation of clathrate-like cages where a perfect network of H-bonds among H$_2$O molecules establishes, as happens for the \openvirg expulsion\closedvirg and \openvirg Sub$_\textsc{CO}$\closedvirg models. As a consequence, CO could not simply substitute water molecules in the ice lattice; 
\item as the concentration increases, CO molecules tend to clusterize and/or rearrange forming homogeneous structures bonded together by attractive dispersive and CO-quadrupole dominated interactions, whose morphology is dictated  by the ice degree of order. Namely, ordered CO layers form on top of a preformed water-ice surface, whereas CO molecules clusterize inside a proper mixed CO-H$_2$O matrix. Our results indicate that small blue- and redshifted peaks could appear in the IR spectrum as a consequence of this \openvirg clusterization\closedvirg process.\\
We hope the present work could be useful to enlighten some aspects of this important astronomical topic, and to provide atomistic details of the CO/H$_2$O ice interactions that could be incorporated in further studies more involved in the problem of the actual structure of interstellar ices.   
\end{enumerate}
\section{Acknowledgments}
This work was financially supported by the Italian MIUR (Ministero dell'Istruzione, dell'Universit\`a e della Ricerca) and from Scuola Normale Superiore (project PRIN 2015, STARS in the CAOS - Simulation Tools for Astrochemical Reactivity and Spectroscopy in the Cyberinfrastructure for Astrochemical Organic Species, cod. 2015F59J3R) and from MINECO (project CTQ2017-89132-P) and from DIUE (Project 2017SGR1320). A.R. is indebted to \openvirg Ram\'on y Cajal\closedvirg program.\\
We acknowledge the funding from the European Research Council (ERC), project DOC (contract 741002).
\bibliographystyle{mnras}


\begin{thebibliography}{}
\makeatletter
\relax
\def\mn@urlcharsother{\let\do\@makeother \do\$\do\&\do\#\do\^\do\_\do\%\do\~}
\def\mn@doi{\begingroup\mn@urlcharsother \@ifnextchar [ {\mn@doi@}
  {\mn@doi@[]}}
\def\mn@doi@[#1]#2{\def\@tempa{#1}\ifx\@tempa\@empty \href
  {http://dx.doi.org/#2} {doi:#2}\else \href {http://dx.doi.org/#2} {#1}\fi
  \endgroup}
\def\mn@eprint#1#2{\mn@eprint@#1:#2::\@nil}
\def\mn@eprint@arXiv#1{\href {http://arxiv.org/abs/#1} {{\tt arXiv:#1}}}
\def\mn@eprint@dblp#1{\href {http://dblp.uni-trier.de/rec/bibtex/#1.xml}
  {dblp:#1}}
\def\mn@eprint@#1:#2:#3:#4\@nil{\def\@tempa {#1}\def\@tempb {#2}\def\@tempc
  {#3}\ifx \@tempc \@empty \let \@tempc \@tempb \let \@tempb \@tempa \fi \ifx
  \@tempb \@empty \def\@tempb {arXiv}\fi \@ifundefined
  {mn@eprint@\@tempb}{\@tempb:\@tempc}{\expandafter \expandafter \csname
  mn@eprint@\@tempb\endcsname \expandafter{\@tempc}}}

\bibitem[\protect\citeauthoryear{Al-Halabi, van Dishoeck  \& Kroes}{Al-Halabi
  et~al.}{2004a}]{JChemPhys_AlHalabi_2004}
Al-Halabi A.,  van Dishoeck E.~F.,   Kroes G.~J.,  2004a, J. Chem. Phys., 120,
  3358

\bibitem[\protect\citeauthoryear{Al-Halabi, Fraser, Kroes  \& van
  Dishoeck}{Al-Halabi et~al.}{2004b}]{AA_AlHalabi_2004}
Al-Halabi A.,  Fraser H.~J.,  Kroes G.~J.,   van Dishoeck E.~F.,  2004b, A\&A,
  442, 777

\bibitem[\protect\citeauthoryear{Allouche, Verlaque  \& Pourcin}{Allouche
  et~al.}{1998}]{JPC_Allouche_1998}
Allouche A.,  Verlaque P.,   Pourcin J.,  1998, J. Phys. Chem. B, 102, 89

\bibitem[\protect\citeauthoryear{B.~Civalleri \& L.~Valenzano}{B.~Civalleri \&
  L.~Valenzano}{2008}]{Grimme_Crystal}
B.~Civalleri C.~M. Z.-W.,  L.~Valenzano P.~U.,  2008, Cryst. Eng. Commun., 10,
  405

\bibitem[\protect\citeauthoryear{Bacmann, Lefloch, Ceccarelli, Castets,
  Steinacker  \& Loinard}{Bacmann et~al.}{2002}]{Bacmann_2002}
Bacmann A.,  Lefloch B.,  Ceccarelli C.,  Castets A.,  Steinacker J.,   Loinard
  L.,  2002, A\&A, 389, L6

\bibitem[\protect\citeauthoryear{Boogert, Gerakines  \& Whittet}{Boogert
  et~al.}{2015}]{ARAA_Boogert_2015}
Boogert A.~C.~A.,  Gerakines P.~A.,   Whittet D.~C.~B.,  2015, ARA\&A, 53, 541

\bibitem[\protect\citeauthoryear{Boys \& Bernardi}{Boys \&
  Bernardi}{1970}]{BSSE}
Boys S.,  Bernardi F.,  1970, Molec. Phys., 19, 553

\bibitem[\protect\citeauthoryear{Casassa \& Pisani}{Casassa \&
  Pisani}{2002}]{JCP_Casassa_2002}
Casassa S.,  Pisani C.,  2002, J. Chem. Phys., 116, 9856

\bibitem[\protect\citeauthoryear{Caselli, Walmsley, Tafalla, Dore  \&
  Myers}{Caselli et~al.}{1999}]{Caselli_CO_depl}
Caselli P.,  Walmsley C.~M.,  Tafalla M.,  Dore L.,   Myers P.~C.,  1999, ApJ,
  523, L165

\bibitem[\protect\citeauthoryear{Ceccarelli, Caselli, Bockel\'ee-Morvan,
  Mousis, Pizzarello, Robert  \& Semenov}{Ceccarelli
  et~al.}{2014}]{Ceccarelli_deut_2014}
Ceccarelli C.,  Caselli P.,  Bockel\'ee-Morvan D.,  Mousis O.,  Pizzarello S.,
  Robert F.,   Semenov D.,  2014, Protostars and Planets VI, University of
  Arizona Press, Tucson, 852

\bibitem[\protect\citeauthoryear{Chetty \& Couling}{Chetty \&
  Couling}{2011}]{JChemPhys_Chetty_2011}
Chetty N.,  Couling V.,  2011, J. Chem. Phys., 134, 164307,164311

\bibitem[\protect\citeauthoryear{Civalleri, D'Arco, Orlando, Saunders  \&
  Dovesi}{Civalleri et~al.}{2001}]{CRYSTAL_optgeom}
Civalleri B.,  D'Arco P.,  Orlando R.,  Saunders V.,   Dovesi R.,  2001, Chem.
  Phys. Lett., 348, 131

\bibitem[\protect\citeauthoryear{Collings, Dever, Fraser  \&
  McCoustra}{Collings et~al.}{2003a}]{Collings_2003a}
Collings M.~P.,  Dever J.~W.,  Fraser H.~J.,   McCoustra M.~R.~S.,  2003a,
  Ap\&SS, 285, 633

\bibitem[\protect\citeauthoryear{Collings, Dever, Fraser  \&
  McCoustra}{Collings et~al.}{2003b}]{Collings_2003b}
Collings M.~P.,  Dever J.~W.,  Fraser H.~J.,   McCoustra M.~R.~S.,  2003b, ApJ,
  585, 1058

\bibitem[\protect\citeauthoryear{Collings, Dever  \& McCoustra}{Collings
  et~al.}{2014}]{Collings_2014}
Collings M.~P.,  Dever J.~W.,   McCoustra M.~R.~S.,  2014, Phys. Chem. Chem.
  Phys., 16, 3479

\bibitem[\protect\citeauthoryear{Dall'Olio, Dovesi  \& Resta}{Dall'Olio
  et~al.}{1997}]{IR_freq}
Dall'Olio S.,  Dovesi R.,   Resta R.,  1997, Phys. Rev. B, 56, 10105

\bibitem[\protect\citeauthoryear{Davidson \& Feller}{Davidson \&
  Feller}{1986}]{BSSE_CP}
Davidson E.,  Feller D.,  1986, Chem. Rev., 86, 681

\bibitem[\protect\citeauthoryear{Davidson, Desando, Cough, Handa, Ratcliffe,
  Ripmeester  \& Tse}{Davidson et~al.}{1987}]{Nature_Davidson_1987}
Davidson D.~W.,  Desando M.~A.,  Cough S.~R.,  Handa Y.~P.,  Ratcliffe C.~I.,
  Ripmeester J.~A.,   Tse J.~S.,  1987, Nature, 328, 418

\bibitem[\protect\citeauthoryear{Devlin}{Devlin}{1992}]{JPhysChem_Devlin_1992}
Devlin J.~P.,  1992, J. Chem. Phys., 96, 6185

\bibitem[\protect\citeauthoryear{Dovesi et~al.,}{Dovesi
  et~al.}{2017}]{CRYSTAL17_manual}
Dovesi R.,  et~al., 2017, {CRYSTAL 2017 User's Manual}

\bibitem[\protect\citeauthoryear{Dovesi et~al.,}{Dovesi
  et~al.}{2018}]{CRYSTAL17_article}
Dovesi R.,  et~al., 2018, WIREs Comput. Mol. Sci.

\bibitem[\protect\citeauthoryear{Dulieu, Amiaud, Congiu, Fillion, Matar, Momeni
   \& Lemaire}{Dulieu et~al.}{2010}]{Dulieu_2010}
Dulieu F.,  Amiaud L.,  Congiu E.,  Fillion J.-H.,  Matar E.,  Momeni A.,
  Lemaire V. P. J.~L.,  2010, A\&A, 512, 5pp

\bibitem[\protect\citeauthoryear{Erba, Casassa, Dovesi, Maschio  \&
  Pisani}{Erba et~al.}{2009}]{JCP_Casassa_2009}
Erba A.,  Casassa S.,  Dovesi R.,  Maschio L.,   Pisani C.,  2009, J. Chem.
  Phys., 130, 074505

\bibitem[\protect\citeauthoryear{Ewing}{Ewing}{1962}]{JChemPhys_Ewing_1962}
Ewing G.~E.,  1962, J. Chem. Phys., 37, 2250

\bibitem[\protect\citeauthoryear{Ewing \& Pimentel}{Ewing \&
  Pimentel}{1961}]{JChemPhys_Ewing_1961}
Ewing G.~E.,  Pimentel G.~C.,  1961, J. Chem. Phys., 35, 925

\bibitem[\protect\citeauthoryear{Favre, Cleeves, Bergin, Qi  \& Blake}{Favre
  et~al.}{2013}]{Favre_2013}
Favre C.,  Cleeves L.~I.,  Bergin E.~A.,  Qi C.,   Blake G.~A.,  2013, ApJ
  Lett., 776, L38

\bibitem[\protect\citeauthoryear{Ferrari, Ugliengo  \& Garrone}{Ferrari
  et~al.}{1996}]{ferrari_CO}
Ferrari A.~M.,  Ugliengo P.,   Garrone E.,  1996, J. Chem.Phys., 105, 4129

\bibitem[\protect\citeauthoryear{Fraser, Collings, Dever  \& McCoustra}{Fraser
  et~al.}{2004}]{Fraser_2004}
Fraser H.~J.,  Collings M.~P.,  Dever J.~W.,   McCoustra M. R.~S.,  2004,
  MNRAS, 353, 59

\bibitem[\protect\citeauthoryear{Frisch et~al.,}{Frisch
  et~al.}{2013}]{Gaussian09}
Frisch M.~J.,  et~al., 2013, GAUSSIAN 09, Revision D.01

\bibitem[\protect\citeauthoryear{Garrod \& Herbst}{Garrod \&
  Herbst}{2006}]{Garrod_2006}
Garrod R.,  Herbst E.,  2006, A\&A, 457, 927

\bibitem[\protect\citeauthoryear{Garrod \& Pauly}{Garrod \&
  Pauly}{2011}]{Garrod_Pauly}
Garrod R.~T.,  Pauly T.,  2011, ApJ, 735, 15

\bibitem[\protect\citeauthoryear{Graham, Imrie  \& Raab}{Graham
  et~al.}{1998}]{MP_Graham_1998}
Graham C.,  Imrie D.~A.,   Raab R.~E.,  1998, Mol. Phys., 93, 49

\bibitem[\protect\citeauthoryear{Grimme}{Grimme}{2006}]{GrimmeD2}
Grimme S.,  2006, J. Comp. Chem., 27, 1787

\bibitem[\protect\citeauthoryear{Grimme}{Grimme}{2010}]{GrimmeD3}
Grimme S.,  2010, J. Chem. Phys., 132, 154104

\bibitem[\protect\citeauthoryear{Hall \& James}{Hall \&
  James}{1976}]{PRB_Hall_1976}
Hall B.~O.,  James H.~M.,  1976, Phys. Rev. B, 13, 3590

\bibitem[\protect\citeauthoryear{Jenniskens, Blake, Wilson  \&
  Pohorille}{Jenniskens et~al.}{1995}]{AstroJ_Jenni_1995}
Jenniskens P.,  Blake D.,  Wilson M.~A.,   Pohorille A.,  1995, ApJ, 455, 389

\bibitem[\protect\citeauthoryear{Karssemeijer, Ioppolo, van Hemert, van~der
  Avoird, Allodi, Blake  \& Cuppen}{Karssemeijer
  et~al.}{2014}]{Karssemeijer_apj_2014}
Karssemeijer L.~J.,  Ioppolo S.,  van Hemert M.~C.,  van~der Avoird A.,  Allodi
  M.~A.,  Blake G.~A.,   Cuppen H.~M.,  2014, ApJ, 781, 1

\bibitem[\protect\citeauthoryear{Kohin}{Kohin}{1960}]{JCP_Kohin_1960}
Kohin B.,  1960, J. Chem. Phys., 33, 882

\bibitem[\protect\citeauthoryear{Kohlmeyer et~al.,}{Kohlmeyer
  et~al.}{2013}]{cp2k_soft}
Kohlmeyer A.,  et~al., 2013, {CP2K}

\bibitem[\protect\citeauthoryear{Lasne, Rosu-Finsen, Cassidy, McCoustra  \&
  Field}{Lasne et~al.}{2015}]{loto1}
Lasne J.,  Rosu-Finsen A.,  Cassidy A.,  McCoustra M. R.~S.,   Field D.,  2015,
  PCCP, 17, 30177

\bibitem[\protect\citeauthoryear{Loveday, Nelmes, Guthrie, Belmonte, Allan,
  Klug, Tse  \& Handa}{Loveday et~al.}{2001}]{Nature_Loveday_2001}
Loveday J.~S.,  Nelmes R.~J.,  Guthrie M.,  Belmonte S.~A.,  Allan D.~R.,  Klug
  D.~D.,  Tse J.~S.,   Handa Y.~P.,  2001, Nature, 410, 661

\bibitem[\protect\citeauthoryear{Madhusudhan, Bitsch, Johansen  \&
  Eriksson}{Madhusudhan et~al.}{2017}]{Madhusudhan_2017}
Madhusudhan N.,  Bitsch B.,  Johansen A.,   Eriksson L.,  2017, MNRAS, 469,
  4102

\bibitem[\protect\citeauthoryear{Manca, Roubin  \& Martin}{Manca
  et~al.}{2000}]{JPL_Manca_2000}
Manca C.,  Roubin P.,   Martin C.,  2000, J. Phys. Lett., 330, 21

\bibitem[\protect\citeauthoryear{Manca, Martin, Allouche  \& Roubin}{Manca
  et~al.}{2001}]{JPC_Allouche_2001}
Manca C.,  Martin C.,  Allouche A.,   Roubin P.,  2001, J. Phys. Chem. B, 105,
  12861

\bibitem[\protect\citeauthoryear{Martin, Manca  \& Roubin}{Martin
  et~al.}{2000}]{SurfSci_Martin_2002}
Martin C.,  Manca C.,   Roubin P.,  2000, Surf. Sci., 502, 280

\bibitem[\protect\citeauthoryear{Mina-Camilde, Manzanares  \&
  Caballero}{Mina-Camilde et~al.}{1996}]{JChemEdu_Mina_1996}
Mina-Camilde N.,  Manzanares I.~C.,   Caballero J.~F.,  1996, J. Chem. Educ.,
  73, 804

\bibitem[\protect\citeauthoryear{Monkhorst \& Pack}{Monkhorst \&
  Pack}{1976}]{Monkhorst}
Monkhorst H.~J.,  Pack J.~D.,  1976, Phys. Rev. B, 13, 5188

\bibitem[\protect\citeauthoryear{Muenter}{Muenter}{1975}]{JMSpec_Muenter_1975}
Muenter J.~S.,  1975, J. Mol. Spectrosc., 55, 490

\bibitem[\protect\citeauthoryear{No\"el, Zicovich-Wilson, Civalleri, D'Arco  \&
  Dovesi}{No\"el et~al.}{2001}]{Berry_phase_crystal}
No\"el Y.,  Zicovich-Wilson C.~M.,  Civalleri B.,  D'Arco P.,   Dovesi R.,
  2001, Phys. Rev. B, 65, 0141/11

\bibitem[\protect\citeauthoryear{Oba, Watanabe, Hama, Kuwahata, Hidaka  \&
  Kouchi}{Oba et~al.}{2012}]{Oba_2012}
Oba Y.,  Watanabe N.,  Hama T.,  Kuwahata K.,  Hidaka H.,   Kouchi A.,  2012,
  Asteroids, Comets, Meteors 2012, Proceedings of the conference held May
  16-20, Japan. LPI Contribution No. 1667

\bibitem[\protect\citeauthoryear{Oberg \& Bergin}{Oberg \&
  Bergin}{2016}]{Oberg_2016}
Oberg K.,  Bergin E.~A.,  2016, ApJ Lett., 831, L19

\bibitem[\protect\citeauthoryear{Oberg, van Dishoeck  \& Linnartz}{Oberg
  et~al.}{2009}]{AA_Oberg_2009}
Oberg K.~I.,  van Dishoeck E.~F.,   Linnartz H.,  2009, A\&A, 496, 281

\bibitem[\protect\citeauthoryear{Palumbo}{Palumbo}{1997}]{JPhysChemA_Palumbo_1997}
Palumbo M.~E.,  1997, J. Chem. Phys. A, 101, 4298

\bibitem[\protect\citeauthoryear{Perdew, Burke  \& Ernzerhof}{Perdew
  et~al.}{1996}]{PBE}
Perdew J.,  Burke K.,   Ernzerhof M.,  1996, Phys. Rev. Lett., 77, 3865

\bibitem[\protect\citeauthoryear{Pisani, Dovesi  \& Roetti}{Pisani
  et~al.}{1988}]{Pisani_book}
Pisani C.,  Dovesi R.,   Roetti C.,  1988, Hartree-Fock Ab Initio Treatment of
  Crystalline Systems.
Springer-Verlag, Berlin, Germany

\bibitem[\protect\citeauthoryear{Pisani, Casassa  \& Ugliengo}{Pisani
  et~al.}{1996}]{CPL_Casassa_1996}
Pisani C.,  Casassa S.,   Ugliengo P.,  1996, Chem. Phys. Lett., 253, 201

\bibitem[\protect\citeauthoryear{Pontoppidan et~al.,}{Pontoppidan
  et~al.}{2003}]{AA_Pontoppidan_2003}
Pontoppidan K.,  et~al., 2003, A\&A, 408, 981

\bibitem[\protect\citeauthoryear{Rosu-Finsen, Lasne, Cassidy, McCoustra  \&
  Field}{Rosu-Finsen et~al.}{2016}]{loto2}
Rosu-Finsen A.,  Lasne J.,  Cassidy A.,  McCoustra M. R.~S.,   Field D.,  2016,
  PCCP, 18, 5159

\bibitem[\protect\citeauthoryear{Sandford, Allamandola, Tielens  \&
  Valero}{Sandford et~al.}{1988}]{AstroJ_Sandford_1988}
Sandford S.~A.,  Allamandola L.~J.,  Tielens A.~G.~G.~M.,   Valero G.~J.,
  1988, ApJ, 329, 498

\bibitem[\protect\citeauthoryear{Sch\"afer, Horn  \& Ahlrichs}{Sch\"afer
  et~al.}{1992}]{Ahlrichs_basiset}
Sch\"afer A.,  Horn H.,   Ahlrichs R.,  1992, J. Chem. Phys., 97, 2571

\bibitem[\protect\citeauthoryear{Schmitt, Greenberg  \& Grim}{Schmitt
  et~al.}{1989}]{AstroPL_Schmitt_1989}
Schmitt B.,  Greenberg J.,   Grim R.~J.~A.,  1989, ApJ Lett., 340, L33

\bibitem[\protect\citeauthoryear{Scuseria}{Scuseria}{1991}]{JCP_Scuseria_1991}
Scuseria G.~E.,  1991, J. Chem. Phys., 94, 6660

\bibitem[\protect\citeauthoryear{Sloan \& Koh}{Sloan \&
  Koh}{2007}]{Clathrate_book}
Sloan E.~D.,  Koh J.~C.,  2007, Clathrate Hydrates of Natural Gases.
CRC Press, Florida, USA

\bibitem[\protect\citeauthoryear{Smith, May  \& Kay}{Smith
  et~al.}{2016}]{Smith_2016}
Smith R.~S.,  May R.~A.,   Kay B.~D.,  2016, J. Phys. Chem. B, 120, 1979

\bibitem[\protect\citeauthoryear{Taquet, Peters, Kahane, Ceccarelli,
  L\'opez-Sepulcre, Toubin, Duflot  \& Wiesenfeld}{Taquet
  et~al.}{2013}]{Taquet_2013}
Taquet V.,  Peters P.~S.,  Kahane C.,  Ceccarelli C.,  L\'opez-Sepulcre A.,
  Toubin C.,  Duflot D.,   Wiesenfeld L.,  2013, aa, 550, A127(1

\bibitem[\protect\citeauthoryear{Tosoni, Pascal, Ugliengo, Orlando, Saunders
  \& Dovesi}{Tosoni et~al.}{2005}]{Tosoni_2005}
Tosoni S.,  Pascal F.,  Ugliengo P.,  Orlando R.,  Saunders V.~R.,   Dovesi R.,
   2005, Mol. Phys., 103, 2549

\bibitem[\protect\citeauthoryear{Vandevondele, Krack, Mohamed, Parrinello,
  Chassaing  \& Hutter}{Vandevondele
  et~al.}{2005}]{CompPhysComm_VandeVondele_2005}
Vandevondele J.,  Krack M.,  Mohamed F.,  Parrinello M.,  Chassaing T.,
  Hutter J.,  2005, Comput. Phys. Commun., 167, 103

\bibitem[\protect\citeauthoryear{Vegard}{Vegard}{1930}]{ZPhys_Vegard_1930}
Vegard L.,  1930, Z. Physik, 61, 185

\bibitem[\protect\citeauthoryear{Zamirri, Corno, Rimola  \& Ugliengo}{Zamirri
  et~al.}{2017}]{ESC_Zamirri_2017}
Zamirri L.,  Corno M.,  Rimola A.,   Ugliengo P.,  2017, ACS Earth and Space
  Chem., 1, 384

\bibitem[\protect\citeauthoryear{Zicovich-Wilson, Pascale, Roetti, Saunders,
  Orlando  \& Dovesi}{Zicovich-Wilson et~al.}{2004}]{Freq_Crystal}
Zicovich-Wilson C.,  Pascale F.,  Roetti C.,  Saunders V.,  Orlando R.,
  Dovesi R.,  2004, J. Comput. Chem., 25, 1873

\makeatother
\end{thebibliography}

%
%
%
\appendix
\section{On-line material}
In the on-line material file we provide: $i$) a detailed discussion of the computational details (basis set, {\it k}-space sampling) and of the equation we used in this work; $ii$) the energetic and vibrational properties of different CO/CO and CO/H$_2$O trimer complexes; $iii$) the resume of all computed IR vibrational wavenumbers and intensities for all models we have developed, as well as a symmetry analysis for \textsc{III}$_{grp1}$ and\textsc{III}$_{grp2}$ CO-containing clathrates; $iv$) the neighbourhood analysis for cage models hosting only one CO molecule {\it per} cage.\\

%
\bsp	
\label{lastpage}
\end{document}